\documentclass[fleqn,usenatbib]{mnras}

\usepackage[T1]{fontenc}

\DeclareRobustCommand{\VAN}[3]{#2}
\let\VANthebibliography\thebibliography
\def\thebibliography{\DeclareRobustCommand{\VAN}[3]{##3}\VANthebibliography}

\usepackage[switch]{lineno}
\usepackage{graphicx}	
\usepackage{amsmath}	
\usepackage{amssymb}	
\usepackage{xspace}
\usepackage{multirow}
\usepackage[normalem]{ulem} 

\newcommand{\fermilat}{\textit{Fermi}-LAT\xspace}

\title[Unveiling  nature of unidentified $\gamma$-ray sources]
{
Unveiling the nature of the unidentified gamma-ray sources
 4FGL J1908.6+0915e, HESS J1907+089/HOTS J1907+091, and 3HWC J1907+085
 in the sky region of the magnetar SGR 1900+14
}

\author[B. Hnatyk et al.]{
B. Hnatyk,$^{1}$\thanks{E-mail: bohdan\_hnatyk@ukr.net}
R. Hnatyk,$^{1}$
V. Zhdanov, $^{1}$
and V. Voitsekhovskyi $^{1}$
\\
$^{1}$ {Astronomical Observatory, Taras Shevchenko National University of Kyiv, 3 Observatorna str. Kyiv, 04053, Ukraine}
}

\date{Accepted XXX. Received YYY; in original form ZZZ}

\pubyear{2022}

\begin{document}


\label{firstpage}
\pagerange{\pageref{firstpage}--\pageref{lastpage}}
\maketitle{}

\begin{abstract}
Supernova remnants (SNRs), star formation regions (SFRs), and pulsar wind nebulae (PWNe)
are prime candidates for Galactic PeVatrons. The nonthermal high-energy
(HE, $\varepsilon>100 \textrm{ MeV}$) and very high-energy 
(VHE, $\varepsilon>100 \textrm{ GeV}$) $\gamma$-ray emission from these sources
should be a promising manifestation  of acceleration processes.  We investigate 
the possibility to  explain the  HE and VHE $\gamma$-ray emission 
from the sky region of the magnetar SGR 1900+14  
 as a signature of cosmic rays accelerated in above mentioned sources.   
To this end, we simulate the $\gamma$-ray emission from the 
extended \fermilat HE source 4FGL J1908.6+0915e, the extended VHE H.E.S.S.
source candidate HOTS J1907+091,
and the point-like HAWC  TeV source 3HWC J1907+085, which are spatially coincident 
with the SNR G42.8+0.6, the  magnetar SGR 1900+14  and the star forming region W49A.
The simulations are performed within the hadronic  and  leptonic
models. We show that   the observed $\gamma$-ray emission from the region of the magnetar 
SGR 1900+14 can, in principle, include contributions of different intensities from all 
three types of  (potentially confused) sources. The considered in detail 
cases of a magnetar-connected but still undetected SNR and a PWN are the  most
promising ones, but with a serious requirement on the energy 
reserve of radiated CR particles - of order of $10^{51}d_{\textrm{10kpc}}^{2}$ erg
for sources at a distance of $d\sim 10$ kpc. Such energy reserve can  be provided by 
the magnetar-related Hypernova and/or magnetar 
wind nebula remnant created by the newborn millisecond magnetar with the large
supply of rotational energy  $E_{\textrm{rot}}\sim 10^{52}\textrm{ erg}$.
  \end{abstract}




\section{Introduction}

 One of the major unsolved problems in modern  astrophysics is the determination of sources and
acceleration mechanism(s) of cosmic rays (CRs) : Galactic ($E_{\textrm{cr}}\lesssim 10^{18}$ eV) and 
extragalactic ($E_{\textrm{cr}}\gtrsim 10^{17}$ eV) ones.
\citep{1990acr..book.....B,2000RvMP...72..689N,2005JPhG...31R..95H,2011ARA&A..49..119K,2013A&ARv..21...70B,2017PhRvL.119s1102N}
Ultra high energy CR 
(UHECR, $E_{\textrm{cr}}>10^{18}$ eV) are believed to be of extragalactic origin, but some
contribution from transient Galactic sources is also possible
\citep{2000RvMP...72..689N,2000ApJ...533L.123B,2003ApJ...589..871A,2011ARA&A..49..119K,2014APh....53..120B,2014JCAP...10..020A,2015JCAP...08..026K,2019FrASS...6...23B}.

In potential Galactic sources of CRs -- 
Supernova (SN) remnants (SNRs), Star formation regions (SFRs), and Pulsar wind nebulae (PWNe) --
powerful shock waves can provide an effective diffusive shock acceleration (DSA)  
of protons, heavier nuclei with a charge $Z>1$, and electrons/positrons up to multi-PeV
energies $E>10^{15}$Z eV
\citep{2005JPhG...31R..95H,2010ApJ...709.1337I,2020SSRv..216...42B,2021MNRAS.504.6096M}.

Promising signatures of acceleration processes are expected to be a nonthermal high-energy 
(HE, $\varepsilon > 100$ MeV) and very high-energy (VHE, $\varepsilon > 100$ GeV) $\gamma$-ray 
emission and neutrino messenger form CR sources' neighbourhoods due to CR nuclei-target 
interstellar matter (ISM) nuclei  inelastic collisions (thereafter $pp$-collisions) with a 
subsequent pion decay (hadronic scenario) and inverse
Compton (IC) scattering of low energy background photons by ultrarelativistic electrons and positrons
(leptonic scenario) 
\citep{1977ApJ...212...60S,2004vhec.book.....A,2013ApJ...774...74F,2018MNRAS.479.3415C,2019scta.book.....C,2021ApJ...919...93F}.
Therefore, the observed spectra and morphology of $\gamma$-ray sources
provide unique information about both the physical processes 
in the sources and the mechanisms of CR acceleration.

Recent observations of sky region of magnetar SGR 1900+14 
reveal a set of new unidentified sources, especially in 
$\gamma$-ray band. They include the extended 4FGL J1908.6+0915e  and three point-like
J1908.7+0812, J1910.0+0904, J1911.0+0905 \fermilat sources
from the 4FGL Catalogue, based on the first eight years of observations from the
\fermilat Gamma-ray Space Telescope \citep{2020ApJS..247...33A}, 
the extended source candidate (hotspot) HOTS J1907+091 detected in the High Energy 
Spectroscopic System (H.E.S.S.) Galactic plane survey (HGPS) \citep{2018A&A...612A...1H},
and   point-like source 3HWC J1907+085/2HWC J1907+084* from the 3HWC catalog of sources with 
energies above several TeV of the High Altitude Water Cherenkov (HAWC)
observatory \citep{2020ApJ...905...76A}. 

One of the potential candidates for these $\gamma$-ray sources  is the magnetar
SGR 1900+14 itself. Magnetars are  young neutron stars with extremely high 
surface magnetic fields $B_{\textrm{s}}\sim 10^{14}-10^{15}$ G 
\citep{1992ApJ...392L...9D,1995MNRAS.275..255T,1996ApJ...473..322T,
2015RPPh...78k6901T,2015SSRv..191..315M,2017ARA&A..55..261K,2021ASSL..461...97E}.
Observed persistent X-ray -- $\gamma$-ray activity of magnetars manifests in forms  of 
repeated hard X-ray -- soft $\gamma$-ray flares (soft gamma repeaters, SGRs)
\citep{1992AcA....42..145P,1995MNRAS.275..255T,1996ApJ...473..322T,1998Natur.393..235K} 
and persistent/transient quiescent super-luminous X-ray emission, together with short 
(milliseconds-seconds) bursts and longer (weeks-months) outbursts (anomalous X-ray pulsars, AXPs) \citep{1995A&A...299L..41V,1996ApJ...473..322T,2003ApJ...588L..93K}.
The total magnetar luminosity considerably exceeds ordinary spin-down energy
losses and is supported mainly
by a decay of magnetar's magnetic energy during a magnetic field evolution 
inside a neutron star or 
reconfigurations of   magnetospheric fields \citep{1995MNRAS.275..255T, 1996ApJ...473..322T}.

Large magnetic fields of magnetars result in a set of distinctive
features of magnetars' evolution in comparison with an ordinary pulsar case.
Dissipation of a dipolar magnetic field energy 
($E_{\textrm{mag}}  \approx B^2_{\textrm{s}} R_{\textrm{NS}}^3/6 \approx 
3\times 10^{47}B_{\textrm{s},15}^2 \textrm{ erg}$ for the radius of neutron star 
$R_{\textrm{NS}}  = 12 \textrm{ km}$) supplies a X-ray -- $\gamma$-ray activity of magnetars,
including spectacular rare giant flares - the  short ($\sim 0.1 \textrm{ s}$) hard X-ray
-- soft $\gamma$-ray bursts with luminosities
$\sim 10^{44} - 10^{47} \textrm{ erg s}^{-1}$ and released energies 
$\sim 10^{44} - 10^{46} \textrm{ erg}$ 
(from SGR 0526–66 on 1979 March 5, from SGR 1900+14
on 1998 August 27, and from SGR 1806–20 on 2004 December 27)\citep{2014ApJS..212....6O}. 
Hereafter the physical quantities are 
expressed by $Q = 10^{n}Q_{\textrm{n}}$ in cgs units, unless specified otherwise.
A sudden release of magnetic energy in over-twisted magnetospheres via solar flare-like 
fast reconnections creates a GRB-like
fireball and  an ultrarelativistic (Lorentz factor $\gamma_\textrm{j} \lesssim 10$) 
outflow \citep{ 1995MNRAS.275..255T, 2013ApJ...774...92P}, observed as radio afterglows 
of the SGR 1900+14 \citep{1999Natur.398..127F} and the SGR 1806–20
\citep{2006ApJ...638..391G} giant flares.

Non-detection of reliable signatures of certainly present magnetar-connected SNR 
and magnetar wind nebula (MWN)  can be explained by  both the large distance  
of $d_{\textrm{mag}} \sim 12.5$ kpc and the 
large absorption in Galactic plane \citep{2014ApJS..212....6O}. Once more,
owing to expected presence of relativistic shock
waves and reconnection processes in  newborn millisecond magnetar winds 
and magnetar giant flares, magnetars are promising accelerators of  UHECRs   
\citep{2000ApJ...533L.123B,2003ApJ...589..871A,2005astro.ph..4452E,2006PASJ...58L...7A,
2010NewA...15..292L, 2011ARA&A..49..119K, 2017A&A...603A..76G, 2019ApJ...878...34F}.
As argued in \cite{2018KPCB...34..167G}, the magnetar SGR 1900+14 is a potential source 
of $E_{\textrm{cr}}>10^{20} \textrm{ eV}$ UHECR triplet
\citep{Sok14} in view of joint data of Telescope Array (TA) \citep{2014ApJ...790L..21A} 
and Auger \citep{2015ApJ...804...15A} detectors. 
Reasonable association of the distant magnetar SGR 1900+14 with the 4 BCE "po star" 
\citep{2002ApJ...569L..43W} suggests a Hypernova type 
of a magnetar-related SN with favourable conditions for UHECR acceleration 
\citep{2011ARA&A..49..119K}. If this is the case, the magnetar SGR 1900+14  should also 
manifest himself as a PeVatron with nonthermal emission in different 
domains -- from radio to TeV-range \citep{2004vhec.book.....A, 2019scta.book.....C}. 

Besides SGR 1900+14, a list of possible source candidates includes two SNRs: SNR G42.8+0.6 
and SNR G 43.3-0.2 from the Catalogue of Galactic Supernova Remnants \citep{2019JApA...40...36G} 
and star forming regions magnetar host stellar cluster Cl 1900+14 and  W49A (Fig.~\ref{fig:SGR_vicinity}).
Possible source confusion should be also taken into account.

In our work we analyse all presently available multi-band observational
data of the magnetar SGR 1900+14 region, including the high-energy
and very high-energy
$\gamma$-ray emission, detected by \fermilat, H.E.S.S. and HAWC. 
We model the observational  data in the frame of an expected hybrid
multi-band emission from  the all potential sources in this sky region: 
magnetar-connected SNR and MWN, SNRs G42.8+0.6 and W49B, SFRs Cl 1900+14 and W49A. 
Source confusion, especially due to TeV halos as a background sources, 
is considered as well. 
As follows from simulations, the most realistic models are connected with hadronic 
or leptonic $\gamma$-ray emission with  the necessary supply of energy
in hadronic or leptonic CR  components of order of 
$10^{51}d_{\textrm{10kpc}}^{-2}$ erg. In the magnetar model  the 
newborn magnetar SGR 1900+14 had to have a millisecond initial 
period in order to transmit $\sim 10^{52}$ erg into the SNR and/or MWN.

We are also reconstructing the evolutionary
track of the magnetar progenitor star from its birth in a young compact 
cluster of massive stars
Cl 1900+14 to its outburst as Hypernova and subsequent  Hypernova remnant (HNR) evolution.

The paper is structured as follows.
In Section \ref{s2} we describe multiwavelength 
 observations of the sky region of the  magnetar SGR 1900+14, from radio  to
 $\gamma$-ray band.  In Section \ref{s3}   expected signatures of potential
 $\gamma$-ray sources in this region-- SNRs, PWNe, and SFRs -- are discussed. 
 Modeling of multiband spectral energy distribution (SED) of observed HE and VHE
 $\gamma$-ray emission from the region of the magnetar SGR 1900+14 are
 presented as well. In Section \ref{s4} we present a comparative analysis of the expected 
 contributions of potential sources to the total observed $\gamma$-ray emission. 
In Section \ref{s5} a comprehensive analysis of Hypernova model as the most
appropriate one is presented. We focus on observational and theoretical arguments
in support of the evolution scenario with the SGR 1900+14
progenitor as a Hypernova/Superluminous Supernova (SLSN). 
The discussion is presented in Section \ref{s6} and conclusions are summarised in Section \ref{s7}.

\section{Multiwavelenght observations of the 
 sky region of the magnetar SGR 1900+14}\label{s2}

\subsection{Sky region of  SGR 1900+14 in the gamma-ray band}\label{s2-1}

Search for persistent and pulsed $\gamma$-ray emission from 20
magnetars (including SGR 1900+14)  
in six years of \fermilat data was  unsuccessful with upper limits
$\sim 10^{-12} - 10^{-11}$ erg cm$^{-2}$ s$^{-1}$ in the $0.1 - 10$ GeV band \citep{2017ApJ...835...30L}. 
At the same time, a positive result of this search was revealing of 
an extended $\gamma$-ray source 
positionally coincident with SGR 1900+14, as well as with the adjacent SNR G042.8+00.6. 
In the {\it Fermi}-LAT Fourth Source Catalogue 
(8-year Source Catalogue 4FGL, v19) 
\citep{2020ApJS..247...33A} this extended source
4FGL J1908.6+0915e with coordinates $\textrm{RA} = 287^{\circ}.16$, $\textrm{Dec} = 9^{\circ}.26$ ($l=43^{\circ}.1249$, 
$b=0^{\circ}.4301$), and radius of $\Theta_{\textrm{ HE}}=0^{\circ}.6$ has an overall 
significance only 4.58$\sigma$ (5.8 $\sigma$ in 4FGL, v.27)
(Fig.~\ref{fig:SGR_vicinity}). Differential photon flux density of 4FGL J1908.6+0915e in the  
50 MeV -- 1 TeV energy range can be approximated by a power-law
\begin{equation}
F_{\textrm{ph}}(\varepsilon)=F_{\textrm{ph}}(\varepsilon_0)(\varepsilon/\varepsilon_0)^{-\Gamma},
\label{eqF}
\end{equation}
with the reference (pivot) energy $\varepsilon_0 =  4.52$ GeV, the normalization
$F_{\textrm{ph}}(\varepsilon_0) = (1.01 \pm 0.19)\times 10^{-13}$ ph cm$^{-2}$s$^{-1}$MeV$^{-1}$
and the photon index $\Gamma = 2.23 \pm 0.098$. As a likely associated source the SNR G42.8+0.6 in this catalogue is noted. 

Very recently \cite{Xin...Guo}
reported  the connection of a piont-like source  4FGL J1910.2+0904c (Fig.~\ref{fig:SGR_vicinity})  
with the $\gamma$-ray emission around W49A -- a massive  star forming region in the Galaxy. 

In the  very high energy (VHE, $\varepsilon > 100$ GeV) $\gamma$-ray domain the H.E.S.S. Galactic Plane Survey (HGPS)
 \citep{2018A&A...612A...1H}
reveals a source candidate (hotspot) HOTS J1907+091 with test statistics (TS) 
above TS = 30 detection threshold only in one (cross-check TS = 43) of two analyses (significance of TS = 18).
Centred at $l= 42^{\circ}.88 \pm 0^{\circ}.08$, $b= 0^{\circ}.69 \pm 0^{\circ}.08$, HOTS J1907+091 has
the measured integral photon flux $F_{\textrm{ph}}(\varepsilon > 1 \textrm{ TeV}) = 4.3\times 10^{-13}$  cm$^{-2}$ s$^{-1}$ and extension 
(radius) of $\Theta_{\textrm{ VHE}}=0^{\circ}.17\pm 0^{\circ}.04$. For purpose of modelling we approximate 
the 1 -- 10 TeV HOTS J1907+091 differential flux 
according to the  \fermilat power-law spectrum  extension
by a power-law spectrum with a  photon index $\Gamma = 2.3 \pm 0.2$ and  a standard deviation
$\delta F = 0.2 F$ (Figs.~\ref{fig:PionIC_PL}--\ref{fig:PWN_TwoPopulations_IC}). Here we took 
into account that average index of known Galactic VHE $\gamma$-ray sources is $\Gamma = 2.3 \pm 0.2$ \citep{2018A&A...612A...1H}.

Two potential counterparts of HOTS J1907+091 are spatially coincident with hotspot:
the magnetar SGR 1900+14 and the SNR G42.8+0.6  \citep{2018A&A...612A...1H}.

Using a blind search methods for  $\gamma$-ray source detection based on image processing
and pattern recognition techniques  \cite{2020APh...12202462R}  have recovered  potentially 
valuable objects in the HGPS significance map. Set of objects around the SGR 1900+14 position 
recovering through an edge detection operator and a Hough circle transform correspond to the 
extended structure of the TeV source  with partial circular symmetry  (Fig.~\ref{fig:SGR_vicinity}).

The closest source to SGR 1900+14  (a separation of $\approx 0^{\circ}.75$) in the 3HWC HAWC Observatory Gamma-Ray
Catalogue \citep{2020ApJ...905...76A} is a TeV point-like source 3HWC J1907+085 with coordinates $\textrm{RA} = 286^{\circ}.79$,
$\textrm{Dec} = 8^{\circ}.57$ ($l=42^{\circ}.35$, $b=0^{\circ}.44$), (all with 
$1\sigma_{\textrm{stat}}=0^{\circ}.09$), a somewhat low
TS = 75.5  peak on an extended background spot (with 2HWC J1907+084* counterpart in the 2HWC Catalogue
\citealt{2017ApJ...843...40A}).

In the 3HWC peak  sensitivity  region ($\sim 10 $ TeV) the best-fit parameters of the 3HWC J1907+085 power-law
$\gamma$-ray spectrum are: the reference (pivot) energy $\varepsilon_0 = 7$ TeV,  the normalization 
$F(\varepsilon_0) = (8.4^{+0.9}_{-1.0}\, ^{+2.5}_{-1.62})\times 10^{-15}$ TeV$^{-1}$ cm$^{-2}$s$^{-1}$ and the  photon 
index $\Gamma = 2.95\pm 0.09\,^{+0.04}_{-0.09}$. A shell-like (necklace) structure 
(with 3HWC J1907+085/2HWC J1907+084* as the brighter knot ($\sigma \sim 5$)), overlapping SGR 1900+14, 
the SNR G42.8+0.6 and H.E.S.S. HOTS J1907+091, are clearly visible in the $\gamma$-ray $\sigma \geq 3$ 
excess map with subtracted $\gamma$-ray contribution from the spatially extended source MGRO J1908+06 
(see Extended Data Fig.~1 in \citealt{2018Natur.562...82A}).

There are no other potential counterparts of 3HWC J1907+085. 
The nearest  ($\sim 0^{\circ}.3$) pulsar 
in the Australia Telescope National Facility Pulsar Catalogue 
(ATNF) of 1509 pulsars \cite{2005AJ....129.1993M} PSR J1908+0839
(period $P=0.185$ s, distance $d = 8.27$ kpc, characteristic age
$\tau_{\textrm{c}} = 1.23$ Myr, spin down power 
$\dot E= 1.5\times10^{34}$ erg s$^{-1}$) is too old and too 
slow rotating in order to support  the necessary TeV luminosity 
of putative PWN \citep{2019MNRAS.485..356A}.

In  recent work \citep{2021arXiv210701425A} the H.E.S.S. Collaboration 
presented a new analysis for the H.E.S.S.
data and  declared that the  the hotspot HOTS J1907+091 is  a real new 
H.E.S.S source HESS J1907+089, namely, it is  is the H.E.S.S. detection
of HAWC J1907+085 source. 
 
To summarise, as the observationally confirmed in the region of the magnetar  SGR1900+14 we can consider only extended HE and VHE $\gamma$-ray  sources: 
the {\it Fermi}-LAT source 4FGL J1908.6+0915e, the H.E.S.S. 
source HESS J1907+089/HOTS J1907+091,
and similar hotspot around HAWC point-like source 
3HWC J1907+085/2HWC J1907+084* (Fig.~\ref{fig:SGR_vicinity}).

\subsection{SGR 1900+14 region in the  radio band}\label{s2-2}

\begin{figure}
    \centering
    \includegraphics[width=1.1 \columnwidth]{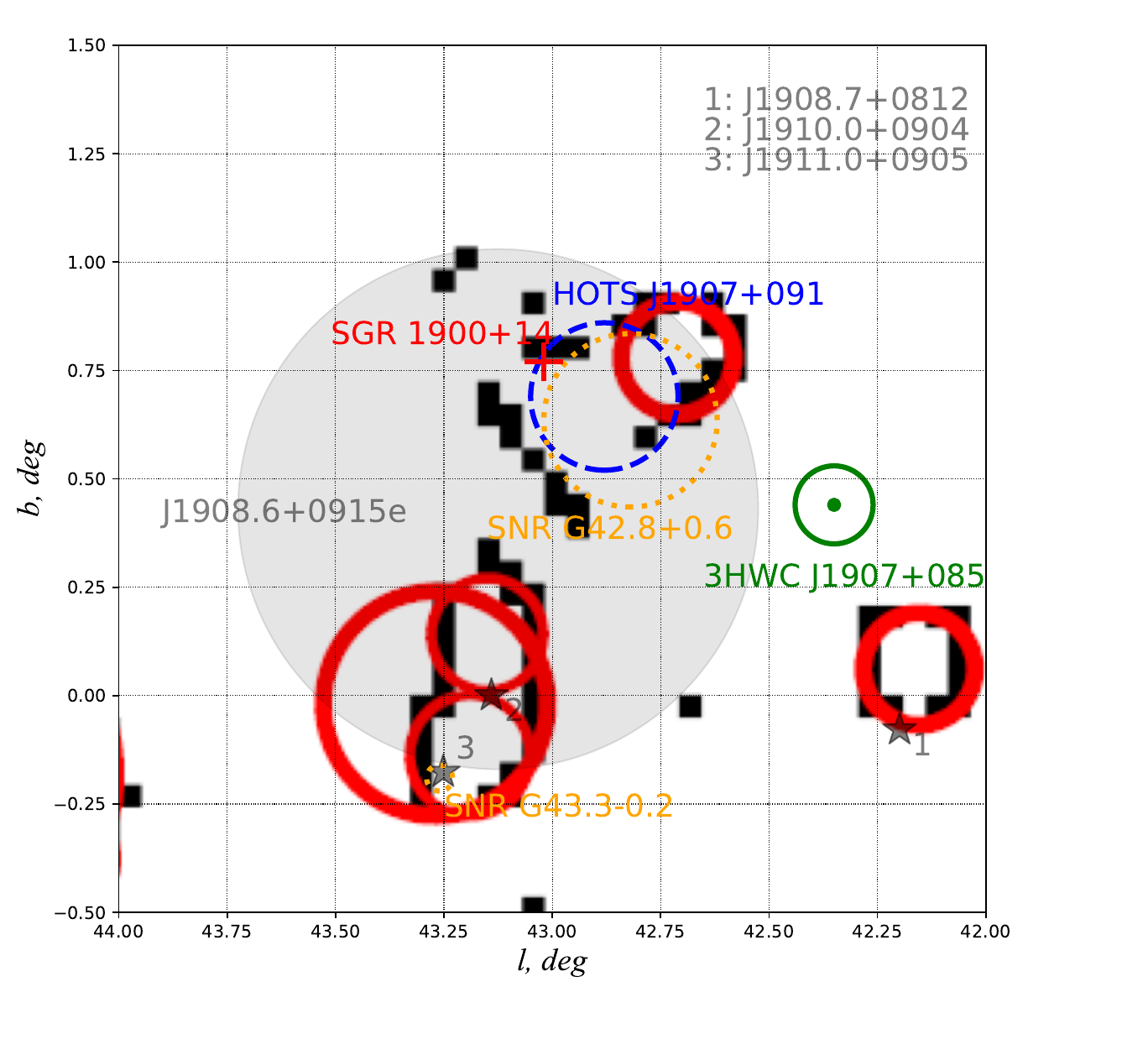}
    \caption{
     Sky region of the magnetar SGR 1900+14. The red cross indicates the SGR 1900+14 position. The
    extended 4FGL source J1908.6+0915e (the grey circle) and three point-like 4FGL  sources (the grey
    stars) are also presented together with the  H.E.S.S. extended source HOTS J1907+091 (the blue ring) 
    and the HAWC point-like source 3HWC J1907+085 (the green bullet with 1$\sigma$ error box). Two SNRs 
    in the field are presented by the orange rings. The star forming region W49A coincides with 
    the star 2. The black filled  squares  correspond to the edge
    mask and the red circles to the objects detected in Hough space in H.E.S.S. data, respectively.
    See text for details.}
    \label{fig:SGR_vicinity}
\end{figure}

The radio  map of SGR 1900+14 region does not display any signatures of the magnetar, 
as well as of a magnetar-connected SNR or a MWN, only $3\sigma$ limits on an extended emission 
of $6.1 \textrm{ mJy arcmin}^{-2}$ (332 MHz) and $6.2 \textrm{ mJy arcmin}^{-2}$ (1.4 GHz) 
were found in \cite{2002ApJ...566..378K}  (Figs.~\ref{fig:PionIC_PL}--\ref{fig:PWN_TwoPopulations_IC}).
 In the proximity to SGR 1900+14 there is only in radio-band detected shell-type SNR  G42.8+0.6
 (of $24^{'}$ size, $l=42^{\circ}.820$, $b=0^{\circ}.635$, spectral flux density
$F_{\varepsilon} = 2.4 \pm 0.6 \textrm{ Jy}$ at 1.420 GHz, $2.0 \pm 0.2 \textrm{ Jy}$ 
at 2.695 GHz,  $1.5 \pm 0.2 \textrm{ Jy}$ at 4.750 GHz, spectral index $\alpha = 0.4$) \citep{1987A&AS...69..403F}.
 The angular separation between the magnetar and the centre (boundary) of the SNR  G42.8+0.6 is of
 $\sim 15^{'}(3^{'})$ (Fig.~\ref{fig:SGR_vicinity}). Young ($\tau_{\textrm{c}}\approx 38 \textrm{ kyr}$) 
 $P=226 \textrm{ ms}$ radio pulsar PSR J1907+0918 ($l=43^{\circ}.024$,  
 $b=0^{\circ}.730$ $d=8.2$ kpc), is a neighbour of SGR 1900+14, the separation between
 the pulsars $\approx 2^{'}$ \citep{2000ApJ...545..385L, 2002ApJ...566..378K}.
Despite the young age of both pulsars, none of them is physically connected with the SNR G42.8+0.6 
in view of  unrealistically high necessary transverse velocity $v_{\textrm{t}}=4\times 10^3 (d_{\textrm{p}}/10 \textrm{ kpc})/(t_{\textrm{p}}/10^4 \textrm{ yr}) \textrm{ km s}^{-1}$ 
for expected distances $d_{\textrm{p}} \sim 5-15 \textrm{ kpc}$ and ages $t_{\textrm{p}}\sim 10^3-4\times 10^4 \textrm{ yr}$ 
of pulsars \citep{1999ApJ...510L.115K, 2000ApJ...545..385L}. Moreover, the estimated 
proper motion of SGR 1900+14 \citep{2012ApJ...761...76T} corresponds to $v_{\textrm{t}} =  (104\pm 24)\times (d_{\textrm{p}}/10 \textrm{ kpc}) \textrm{ km s}^{-1}$
and  to the approach to the SNR with position angle $\textrm{PA}=254^{\circ} \pm 10^{\circ}$). Recent estimate of the extinction-based 
distance to the SNR  G42.8+0.6 $d_{\textrm{ext}}=4.24 \pm 0.93$ pc \citep{2020A&A...639A..72W}  also excludes its
physical connection with SGR 1900+14. 

The second, more distant ($\sim 1^{\circ}$)  from the magnetar, shell-type SNR G43.3-0.2  
(W49B), of $4^{'}\times 3^{'}$ size  at distance $d=11.3$ kpc, $l=43^{\circ}.275$, $b=-0^{\circ}.190$, 
spectral flux density  $F_{\varepsilon} = 38 \textrm{ Jy}$ at 1 GHz, 
spectral index $\alpha = 0.46$) \citep{2019JApA...40...36G}
is detected also in X-ray and $\gamma$-ray range  as a young 
(age  $t\approx$ 6 kyr)  compact (radius $R\approx 8$ pc) SNR --  effective cosmic
ray accelerator with the proton CR energy $W_{\textrm{cr},p}\approx 2\times 10^{49}$ erg), evolving in a  dense
( the  number density $n_{\textrm{mc}}\approx 650$ cm$^{-3}$) molecular cloud (see \cite{2021ApJ...919..123S}
and references therein). Due to large separation (more than 1 kpc) W49B  does not belong to the
SGR 1900+14 neighborhood. 

\subsection{SGR 1900+14 region in the  optical/infrared band}\label{s2-3}

Initial searches for a SGR 1900+14 counterpart in the optical/infrared (IR) region  were unsuccessful: 
HST/STIS, and  Keck J/Ks-band images placed  only upper limits of 
$m_{\textrm{50CCD}}\gtrsim 29.0$ mag, $J\gtrsim 22.8$ mag, $K_s\gtrsim 20.8$ mag \citep{2002ApJ...566..378K}.
Only in \cite{2008A&A...482..607T} a newly detected variable "object \# 7" with $K_s \approx 19.7$ mag and 
$\Delta K_s=0.47 \pm 0.11$ mag,  observed with  ESO-VLT/NACO in the error circle of the SGR 1900+14 radio position,
was proposed  as a tentative IR candidate counterpart to SGR 1900+14. Later on, "Star 7" was confirmed by Keck 2
LGS-AO/NIRC2 observation as the IR counterpart \citep{2012ApJ...761...76T}. 

\cite{2012ApJ...761...76T}  confirmed claimed in 
\cite{1996ApJ...468..225V,2000ApJ...533L..17V} physical 
connection of the magnetar with  the  young 
compact of radius $R_{\textrm{cl}}\approx 0.4-0.6$ pc  embedded cluster of 
massive stars  Cl 1900+14,  
aka SGR 1900+14 Star Cluster  \citep{2013A&A...560A..76M}, with two luminosity-dominated
M5 Red Supergiant (RSG) and a few Blue Supergiant (BSG) 
stars and  improved in  \cite{2009ApJ...707..844D} distance 
$d_{\textrm{cl}}= 12.5\pm 1.7 \textrm{ kpc}$ and  
extinction   $A_V=12.9 \pm 0.5$ mag. In view of the distance, having the  transverse
velocity $v_{\textrm{t}} =130\pm 30 \textrm{ km s}^{-1}$, 
 it is possible to estimate the age of the magnetar which  needs 
$6\pm 1.8\, (3 \pm 0.9) \textrm{ kyr}$ to escape from its birthplace
at the centre (edge) of the cluster. 

\begin{figure}
    \centering
    \includegraphics[width=0.95\columnwidth]{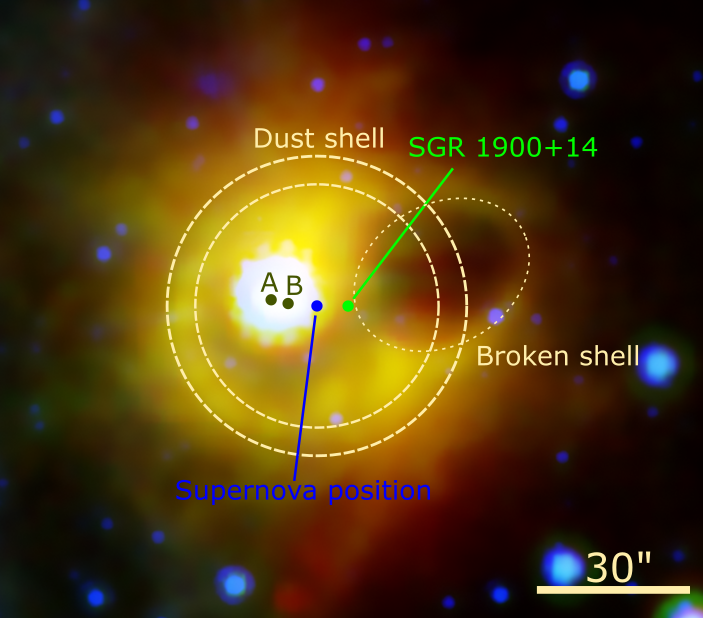}
    \caption{Spitzer IR map of SGR 1900+14 environment with colour coding: 
    blue (8 $\mu$m), green (16 $\mu$m), 
    and red (24$\mu$m) (courtesy NASA/JPL/Caltech/S. Wachter (Caltech-SSC)).
    We indicated with the green bullet 
    the present magnetar position and with the black bullets A and B -- 
    the positions of two RSGs 
    at the centre of the open star cluster Cl 1900+14/SGR 1900+14. 
    In our model, the blue bullet in the centre of  
    the observed dust shell (the yellow dashed ring) indicates the 
    Supernova position 2 kyr ago, whereas the 
    broken shell (the yellow dotted ellipse) has been created by the
    anisotropic giant flare on 1998 August 27.     See text for details.}
    \label{fig:IR_ring}
\end{figure}

An additional confirmation of a physical connection of the magnetar
with the cluster is based  on the IR 
elliptical  shell around SGR 1900+14 (Fig.~\ref{fig:IR_ring}),  discovered by 
\cite{2008Natur.453..626W} in Spitzer $16 \:\mu \textrm{m}$ and 
$24 \:\mu \textrm{m}$ images.  \cite{2008Natur.453..626W} suggested that
this shell is a signature of a
dust-free cavity produced 
by the giant flare of SGR 1900+14 in the stellar cluster-connected 
dusty gas on 1998 August 27. Observed shell dust 
temperature of 80 -- 120 K and IR flux  of $1.2\pm 0.2 \textrm{ Jy}$ and 
$0.4\pm 0.1 \textrm{ Jy}$ at 
$24 \:\mu \textrm{m}$ and $16 \:\mu \textrm{m}$, respectively, together with
the absence of ring signatures
at optical, near-IR, radio- or X-ray wavelengths, was explained in a model 
of a dust heating by the embedded 
star cluster without the possible presence of a shock wave from the SN 
explosion. The 3D dust radiative transfer modelling  
\citep{2017ApJ...837....9N} reproduces observational data in a model 
of a cavity in a circumcluster medium with a sharp inner boundary and a
molecular cloud-like extinction, 
with two illuminating RSG stars inside the cavity. The necessary total
mass of the radiating dust is $M_{\textrm{dust}} \sim 2{\textrm{M}}_{\odot}$ 
and the  cloud gas number density 
$n_\textrm{mc}\sim 10^3 \textrm{ cm}^{-3}$ for a dust-to gas ratio of 0.00619. 
Resulting high gas number density at the cavity boundary leads to the  total mass of 
gas inside the IR shell of radius $\sim 1.5$ pc and  of volume 
$\sim 13 \textrm{ pc}^3$ of order of  $ 400 {\textrm{M}}_{\odot}$ and
the total mass of gas outside the cavity 
$\sim 320 {\textrm{M}}_{\odot}$, or the total mass of a putative molecular 
cloud of radius $\sim 2$ pc
$M_{\textrm{mc}}\gtrsim 10^3 {\textrm{M}}_{\odot}$, but such a molecular 
cloud is still not observed.
In the  Catalogue of of 695 embedded and open stellar clusters in the inner Galaxy
VizieR J/A+A/560/A76 \citep{2013A&A...560A..76M} 
the cluster Cl1900+14/SGR 1900+14 is classified as an "OC2: 
open cluster without gas and without ATLASGAL survey counterpart".
In a complete sample of $\sim$ 8000 
$M_{\textrm{mc}} \gtrsim 10^3 {\textrm{M}}_{\odot}$ dense clumps located in the Galactic
disc ($5^{\circ} < l < 60^{\circ}$) 
ATLASGAL \citep{2018MNRAS.473.1059U} an angular separation with the closest clump is
of $0^{\circ}.8$ (G043.141-00.01),
whereas in the all-Galaxy CO survey 
($M_{\textrm{mc}} \gtrsim 3\times 10^3 {\textrm{M}}_{\odot}$ in the outer Galaxy)
\citep{2016ApJ...822...52R} the closest molecular cloud with 
$l=43^{\circ}.553$, $b=0^{\circ}.567$ is of $0^{\circ}.57$ away. 

Therefore in Section \ref{s5-3} we propose and substantiate a new model of observed 
IR emission  from the relics
(debris) of  dusty magnetar-related SN ejecta (Fig.~\ref{fig:IR_ring}).

\subsection{SGR 1900+14  region in the  X-ray band}\label{s2-4}

In the X-ray domain, after the giant flare in 1998 and two re-brightening in 
2001 and 2006, SGR 1900+14 returned to
minimum  quiescent flux level with the typical for magnetars two-component 
spectrum: a black body component with
temperature of $k_{\textrm{B}}T= 0.52$ keV and a hard power-law component 
with photon index $\Gamma = 1.21$.
For a column density of $N_\textrm{H} = 1.9\times 10^{22}$ cm$^{-2}$, 
the quiescent  bolometric luminosity 
is of $L_{\textrm{bol}} = 5.6\times 10^{35}$ erg s$^{-1}$
\citep{2015RPPh...78k6901T, 2017ApJS..231....8E, 2018MNRAS.474..961C,2019PASJ...71...90T}. 
Recent analysis of XMM-Newton
and NuSTAR data \citep{2019PASJ...71...90T} has revealed a monotonic decrease
in the 1 -- 10 keV flux and the spin-down rate $\dot P$ in 2006-2016. 
Newly obtained values $P=5.22669(3)$ s, $\dot P=3.3\times 10^{-11}$ s s$^{-1}$ for 2016 
correspond to   magnetar magnetic field $B_{\textrm{s}} = 4.3\times 10^{14}$ G and the characteristic age
$\tau_{\textrm{c}}\approx 2.4\textrm{ kyr}$ (cf. the  old value 
$\tau_{\textrm{c}}\approx 700 \textrm{ yr}$). No signatures of an  adjacent
SNR or MWN were detected yet in the  X-ray band.

We adopted the intrinsic (unabsorbed) 2 -- 10 keV flux of  SGR 1900+14 $F_{\textrm{{2-10\,keV}}}=1.8\times 10^{-12}$ erg
cm$^{-2}$s$^{-1}$ = $1.1\times 10^{-9}$ GeV cm$^{-2}$s$^{-1}$ 
\citep{2019PASJ...71...90T} as an upper limit for expected SNR/MWN  2-10 keV flux.  

\section{Expected signatures of SNRs, SFRs and PWNe in the SGR 1900+14 region}\label{s3}

Detected in the SGR 1900+14 region   HE and VHE $\gamma$-ray sources: extended ones 4FGL
J1908.6+0915e and  HOTS J1907+091, as well as point-like peaks on enhanced  background 
3HWCJ1907+085/2HWCJ1907+084* and  HESS J1907+089 correspond to three potential
acceleration places of relativistic hadrons and leptons: SNRs, SFRs and PWNe.
Due to presence of powerful shock waves with effective diffusive shock 
acceleration of CRs they are prominent Galactic PeVatrons -- accelerators of protons,
heavier  nuclei and  electrons/positrons up to PeV-energies. In observations, 
they are manifested by nonthermal radiation, especially, by HE and VHE $\gamma$-ray  emission.

\subsection{SNRs, SFRs and PWNe as CR accelerators}\label{s3-1}

\subsubsection{SNRs as CR accelerators}\label{s3-1-1}
There are at least two SNRs in the SGR 1900+14 region: SNR G042.8+0.6 
and the SNR of a SN that gave birth to a magnetar but has not yet 
been identified (Fig. \ref{fig:SGR_vicinity}). Indeed, 
magnetar SGR 1900+14, like other neutron stars, is a compact remnant of 
core collapse SN explosion.

In a case of  an ordinary SN explosion a SN ejecta with a typical kinetic energy 
$W_{\textrm{ej}}\sim$ 1 Bethe (B) $= 10^{51}$ erg (from 0.09 to 2.3 B for 9-27 
M$_\odot$ progenitors) interacts with
circumstellar/interstellar medium (ISM), creating shock-bounded expanding bubble 
of swept-up ISM gas -- SNR -- with   energy reserve of 
$W_{\textrm{SNR}}\sim W_{\textrm{ej}} \sim 10^{51}$ erg.
A newborn neutron star -- typical radio pulsar with a magnetic field 
$B_{\textrm{s}}\sim 10^{12}$ G, a rotational period $P_i\sim 0.01-0.1$ s 
and a rotational energy 
$E_{\textrm{rot}}\sim 10^{48}-10^{50}$ erg, creates a termination shock (TS)-bounded 
PWN inside the SNR  with an energy reserve of 
$W_{\textrm{PWN}}\lesssim E_{\textrm{rot}} \lesssim 10^{50}$ erg
(see \cite{2021Natur.589...29B} and references therein). 

SNRs  are effective accelerators of Galactic CRs (of both hadronic 
(protons and nuclei) and leptonic (electrons) components) via DSA
of charged ISM particles at SNR shocks.
 
The expected spectra of shock-accelerated nuclei (mainly, protons, 
$\textrm{i}=p$) and electrons ($\textrm{i}=e$) are of power-law (PL) 
or   exponential cut-off power-law (ECPL) type
\begin{equation} \label{eq:20}
N_{\textrm{i}}(E_{\textrm{i}})=N_{\textrm{0,i}}
    (E_{\textrm{i}}/E_{\textrm{0,i}})^{-\Gamma_{\textrm{cr,i}}} 
    \exp(-E_{\textrm{i}} /E_{\textrm{cut,i}}),
\end{equation}
where $E_{0,\textrm{i}}$ is the CR reference energy, $N_{\textrm{0,i}}$ 
is the normalisation, $N_{\textrm{0},e}(E)=K_{ep}N_{\textrm{0},p}(E)$, $K_{ep}$ 
is the electron-to-proton energy fraction, $\Gamma_{\textrm{cr,i}}$ and 
$E_{\textrm{cut,i}}$ are the spectral index and the cut-off energy of i-th component,
respectively. Classical value for spectral index in Fermi-I nonrelativistic  
shock acceleration is
$\Gamma_{\textrm{cr,i}}=2$, in case of relativistic  parallel shocks 
$\Gamma_{\textrm{cr,i}}=2.2$ \citep{1999JPhG...25R.163K,2000ApJ...542..235K}.
Observations of Galactic CR sources, e.g. radio-sources with spectral  energy 
flux $F_{\varepsilon}\propto \varepsilon^{-\alpha}$
with $\alpha= \Gamma+1= (\Gamma_{\textrm{cr},e}-1)/2 \approx (0.6 - 0.8)$ and 
diffuse Galactic CR backgroud correspond to $\Gamma_{\textrm{cr,i}}=2.2-2.5$
\citep{1971NASSP.249.....S,1990acr..book.....B,2005JPhG...31R..95H,
2010A&A...524A..51D,2012AdSpR..49.1313F,2017PhRvD..95h3007Y,2017SSRv..207..319P,
2021PhRvD.103h3010E}.  
 
The total energy of CR $W_{\textrm{cr,i}}$ ($\textrm{i} = p, e$) is 
\begin{equation} \label{eq:wtot}
W_{\textrm{cr,i}}=\int_{E_{\textrm{min,i}}}^{E_{\textrm{max,i}}}
EN_{\textrm{0,i}}
(E/E_{\textrm{0,i}})^{-\Gamma_{\textrm{cr,i}}}
\exp(-E/E_{\textrm{cut,i}})dE,
\end{equation}
and accelerated at SNR shock CR components include hadrons with total energy
$W_{\textrm{cr},p}= \eta_{\textrm{cr}} W_{\textrm{SNR}}$ and leptons 
with total energy $W_{\textrm{cr},e}=K_{ep}W_{\textrm{cr},p}$,
where $\eta_{\textrm{cr}}\sim 0.1$ is the efficiency of CR acceleration.

In typical SNR with $W_{\textrm{SNR}}\sim 10^{51} \textrm{ erg}$ we expect
$W_{\textrm{cr},p}\sim 0.1\eta_{p,-1}W_{\textrm{SNR}} 
\sim 10^{50}\eta_{p,-1} \textrm{ erg}$, 
$W_{\textrm{cr},e} \sim 0.01K_{ep,-2} W_{\textrm{cr},p}\sim 10^{48}
K_{ep,-2}\textrm{ erg}$ for young SNR (age $t_{\textrm{SNR}}\lesssim 10^4$ yr)
with radius of $R_{\textrm{SNR}}(t\sim 10^3 \textrm{ yr}) \lesssim 10 \textrm{ pc}$
\citep{1999ApJS..120..299T,2008ARA&A..46...89R}.
Hereafter the physical quantities are 
expressed by $Q = 10^{n}Q_{\textrm{n}}$ in cgs units, unless specified otherwise.

Maximum energy of the SNR diffusive shock-accelerated CRs with charge $Ze$, velocity
$V_{\textrm{sh}}=\beta_{\textrm{sh}}\cdot c$ and   shock-compressed ISM magnetic field 
$B_{\textrm{sh}}=\sqrt{11}B_{\textrm{ISM}}$ is   \citep{2005JPhG...31R..95H}: 
\begin{equation} \label{eq:19}
E_{\textrm{max}}\approx  Ze \beta_{\textrm{sh}} R_{\textrm{SNR}} B_{\textrm{sh}}\approx \\
3\times 10^2 Z\beta_{\textrm{sh,-2}}R_{\textrm{SNR,19}}
B_{\textrm{sh,-5}} \textrm{ TeV},
\end{equation}
or $E_{\textrm{max}}\approx 0.9Z  \textrm{ PeV}$  in our case.

Accelerated CRs, accumulated in a downstream region, are subject 
to adiabatic losses due to nonzero divergence of SNR plasma 
expansion velocity $\bar{v}$ \citep{2012MNRAS.427..415M}:
\begin{equation}
\label{tad}
{\dot E}_{\textrm{i,ad}}=-\frac 13(\nabla \cdot \bar{v})E_{\textrm{i}}.
\end{equation}
In a spherically symmetric SNR 
$E_{\textrm{i}}\propto R_{\textrm{SNR}}^{-1}\propto t_{\textrm{SNR}}^{-0.4}$,
and the adiabatic loss time 
$t_{\textrm{ad}}=(E_{\textrm{i}}/{\dot E_{\textrm{i}}})_{\textrm{ad}}= 2.5t_{\textrm{SNR}}$
exceeds the SNR age. At the same time, hadronic CRs diffusively escaping into upstream
region with density enhancement $n_{\textrm{ISM}}\gtrsim 1$ cm$^{-3}$ are noticeably losing 
energy in collisions with ISM protons and nuclei. For proton-proton collision 
cross-section $\sigma_{pp}=40$ mb and the coefficient of inelasticity - part of
proton energy lost in interaction - $\kappa\approx 0.5$ the cooling time is 
$t_{pp}=(\kappa n_{\textrm{ISM}}\sigma_{pp}c)^{-1}
\approx 1.7\times 10^{15}n_{\textrm{ISM}}$ s \citep{2004vhec.book.....A}. 
Produced in collisions 
secondary pions fastly decay into leptons, in particular, hadronic mechanism of 
$\gamma$-ray generation is connected with  neutral pion decay
$pp \to \pi^{\circ} \to \gamma \gamma$. 
In the hadronic $E_{p}\gtrsim  10$ GeV 
scenario  CR protons with energy $E_{p}\gtrsim  10$ GeV produce $\gamma$-ray 
photons with energy $\varepsilon\approx 0.1 E_{p}$ and CR protons with  
power-law spectrum (\ref{eq:20}) produce power-law spectrum of 
$\gamma$-ray photons (\ref{eqF}) with the spectral photon index
$\Gamma=\Gamma_{\textrm{cr},p}$. 
Time scale for $\gamma$-ray production  is  
$t_{pp,\gamma}=(\kappa_{\pi^0} n_{\textrm{ISM}}\sigma_{pp}c)^{-1}
\approx 5 \times 10^{15}n_{\textrm{ISM}}$ s, 
where $\kappa_{\pi^0}\approx 0.17$ is  the fraction of proton kinetic 
energy transferred to $\pi^0$
\citep{1971NASSP.249.....S,1990acr..book.....B,2018A&A...612A...3H}.

Ultra high energy protons with Lorentz factor $\gamma_p$ such that 
energy of background photon $\varepsilon$ in proton rest frame 
$\varepsilon' =\gamma_p\varepsilon > 145 \textrm{ MeV}$ produce
mesons (pions) in  proton-photon collisions 
$p+\gamma\to p+\pi^0$, $p+\gamma \to n+\pi^{+}$. For CMB photons with
$\varepsilon_{\textrm{CMB}}\approx 3 k_B T_{\textrm{CMB}}\approx 7\times10^{-4}$ eV,
where $k_B$ is the Boltzmann constant, 
$\gamma_p \geq 10^{11}$. Cooling time for photopion production  is 
$t_{p\gamma}=(\kappa_{p\gamma} n_{\gamma}\sigma_{p\gamma}c)^{-1}
\approx 1.2 \times 10^{15}(n_{\gamma}/n_{\textrm{CMB}})^{-1}$ s,
where $\kappa_{p\gamma}\approx 0.2$ is the coefficient of inelasticity  and 
$\sigma_{p\gamma}\approx 340$ $\mu$b is the cross-section for the 
photopion production \citep{1971NASSP.249.....S,1990acr..book.....B}.
For Galactic CRs accelerated at SNR shocks a  photpion production 
is negligible.  

Energy losses of protons due to nonthermal (synchrotron and IC)
radiation are $(m_p/m_e)^4 \approx 1.1\times 10^{13}$ times smaller than the 
losses of electrons of the same energy (see detail below).

Observable spectra of hadronic $\gamma$-ray emission are determined by
parameters of hadronic CR 
spectra (\ref{eq:20}) and number densities of target ISM particles $n_{\textrm{ISM}}$
\citep{2016ApJS..224....8A,2018MNRAS.479.3415C,2018A&A...612A...1H}.

\subsubsection{SFRs  as CR accelerators}\label{s3-1-2}

Galactic SFRs,  that include stellar
clusters, OB associations, and superbubbles, are also promising  sources
of Galactic CRs and potential Galactic PeVatrons 
\citep[][and references therein]
{2020SSRv..216...42B,2021MNRAS.504.6096M,2021Univ....7..324C}.

In the SGR 1900+14 region there are two SFRs:  magnetar-connected star
cluster Cl 1900+14  and a W49 complex that consists of a star forming 
region W49A  and a SNR W49B (the last one outside the region considered) (Fig. \ref{fig:SGR_vicinity}). 

In SFRs a common action of powerful winds of massive stars and  SN ejectas 
leads to formation of  a set of strong individual and large-scale termination shocks.
Analogously to the considered above  SNR case, DSA is
effective here and provides power-law type of CR spectra with spectral indexes
$\Gamma_{\textrm{cr}}=2-2.3$. Once more,  a presence of multiple shocks increases 
the efficiency   of  CR acceleration  and their maximum energies
\citep{2020SSRv..216...42B}. VHE $\gamma$-ray emission have been detected from some  
SFRs  including magnetar SGR 1806-20-connected  star cluster
Cl*1806-20 \citep{2018A&A...612A...2H,2018A&A...612A..11H,2021MNRAS.504.6096M}.

\subsubsection{PWNe  as CR accelerators}\label{s3-1-3}

PWNe  are also effective accelerators of CRs, but mainly  of
their leptonic (electron and positron)  components as dominant
pulsar wind constituents via DSA at the  relativistic TS \citep{2006ARA&A..44...17G,
2011MNRAS.410..381B, 2014IJMPS..2860160A,
2015SSRv..191..391K,2017ApJ...845..139B,
2017SSRv..207..175R, 2018A&A...612A...2H,  
2018A&A...609A.110Z,2020arXiv200104442A},
as well as via  a driven magnetic reconnection (MR)  and  a resonant
absorption of ion-cyclotron waves \citep{2014IJMPS..2860160A,2020arXiv200104442A}.

Complex nature of lepton CR  acceleration in PWN case does not allow 
to reliably determine CR spectral characteristics.    
The present relativistic leptonic component of CRs 
in PWNe is the result of the time-dependent TS/MR acceleration and 
cooling/escaping/advection 
processes \citep[][and references therein]{2018A&A...612A...2H, 2018A&A...609A.110Z}. 
The observed SED in PWNe is the sum of leptonic contributions from different 
injection epochs and space localisation. In this case a typical leptonic 
exponential cut-off broken power-law (ECBPL)  spectrum 
\begin{equation} \label{eq:23}
N_{e}(E_{e})=
 \begin{cases}
 N_{\textrm{0},e}(E_{e}/E_{\textrm{0},e})^{-\Gamma_{\textrm{1,cr},e}}, E_{e}<E_{\textrm{br},e} \\
\begin{split}
N_{\textrm{0},e}(E_{\textrm{br},e}
/E_{\textrm{0},e})^{(\Gamma_{\textrm{2,cr},e}-\Gamma_{\textrm{1,cr},e})}
(E_{e}/E_{\textrm{0},e})^{-\Gamma_{\textrm{2,cr},e}}\\
\times \exp(-E_{e}/E_{\textrm{cut},e}), E_{e}\geq E_{\textrm{br},e}
\end{split}
\end{cases}
\end{equation}
with characteristic break energy $E_{\textrm{br},e}$ and two PL spectra with  $\Gamma_{\textrm{1,cr},e}< 2$ for $E_{e} \leq E_{\textrm{br},e}$ and 
$\Gamma_{\textrm{2,cr},e} > 2$ for $E_{e} \geq E_{\textrm{br},e}$ is widely used in PWN  modelling  \citep[e.g.][]{2014JHEAp...1...31T,2017ApJ...838..142I}.

Acceleration of leptonic components in young ($t\lesssim t_{\textrm{sd,i}}$) PWNe  is very
efficient. Considerable part $\eta_{e}$
of spin-down luminosity $L_{\textrm{sd}}$ is converted into CR lepton energy
$W_{\textrm{cr},e}$: 
$\eta_{e}=\dot W_{\textrm{cr},e}/L_{\textrm{sd}}\lesssim 1$ , but energy losses of accelerated 
leptons result in an actual (derived from observations) value of
$W_{\textrm{cr},e}\lesssim 10^{49}$ erg \citep[][and references therein]{2018A&A...612A...2H}.

Maximum energy of  CR leptons, accelerated at relativistic TS with  the 
TS velocity $\beta_{\textrm{TS}}=V_{\textrm{TS}}/c\approx 1$,  the radius of PWN
$R_{\textrm{PWN}}\sim 10^{19}$ cm,  and the magnetic field inside it 
$B_{\textrm{PWN}}\sim 100$ $\mu \textrm{G}$ is considerably smaller than $\sim 300$ PeV, 
predicted by (\ref{eq:20}) due to their severe energy losses inside PWN. 

Relativistic electrons
in  the SNR/SFR case and electrons/positrons the PWN case (hereafter 
we use the term electrons to the both electrons and positrons in the last 
case) interact with  magnetic fields of energy density $w_B=B^2/8\pi$ and with
components of background radiation with energy density $w_{\textrm{rad,i}}$: 
cosmic microwave background (i = CMB), infrared (i = IR), 
star light (i = SL) radiation, producing
synchrotron (in  the radio -- X-ray band) and IC -- 
synchrotron self-Compton (SSC) (in the GeV -- TeV band)  emission.
\citep{2006ARA&A..44...17G,2015SSRv..191..391K,
 2018A&A...609A.110Z,2020arXiv200104442A}. 
 
For a relativistic electron with the Lorentz factor $\gamma_e>>1$ 
and energy $E_e=\gamma_em_ec^2$ the rate of energy losses is
\begin{equation} \label{EnLos}
P(E_e)=-\frac{dE_e}{dt}=\frac{4}{3}\sigma_T c(w_B+\Sigma_\textrm{i} w_{\textrm{rad,i}}f_{\textrm{KN,i}}(E_e,T_i))\gamma_e^2,
\end{equation}
where $\sigma_T$ is the Thomson cross-section and 
\begin{equation} \label{fkn}
f_{\textrm{KN,i}} = \frac{45K^2/64\pi^2}{45K^2/64\pi^2+ \gamma_e^2},
\end{equation}
where $K=m_ec^2/k_B T_\textrm{i}$   and $T_\textrm{i}$ is the 
temperature of the i-th component of background radiation (i = CMB, IR, SL),
describes transition to Klein-Nishina scattering 
\citep[][and references therein]{2020PhRvL.125e1101E}.
From (\ref{EnLos}) it  follows that electron energy loss time 
in the Thomson regime  for synchrotron emission is 
\begin{equation} \label{tsyn}
t_{\textrm{syn}}(\gamma_e)=\frac{E_e}{P_{\textrm{syn}}(E_e)} 
=\frac{6\pi m_e c}{\sigma_T B^2\gamma_e}=
2.4\times10^{1}B_{-5}^{-2}\gamma_{e,7}^{-1} \textrm{ kyr},
\end{equation}
whereas the energy loss time for IC emission is
\begin{equation} \label{tic}
t_{\textrm{IC,i}}(\gamma_e) = \frac{3 m_e c}{4\sigma_T w_{\textrm{rad,i}}\gamma_e} 
=2.6\times 10^{2}\frac{w_{\textrm{rad,CMB}}}{w_{\textrm{rad,i}}}\gamma_{e,7}^{-1} \textrm{ kyr}.
\end{equation}
Here $w_{\textrm{rad,CMB}}=0.26$ eV cm$^{-3}$ and 
$B_{\textrm{CMB}}=(8\pi w_{\textrm{rad,CMB}})^{1/2}=3.2\, \mu \textrm{G}$.

Advecting in downstream flow CR electrons suffer also  from adiabatic losses with adiabatic time loss (\ref{tad}) determined by PWN radius $R_{\textrm{PWN}}$ and flow velocity $V_{\textrm{PWN}}$: 
$t_{\textrm{ad, PWN}}=R_{\textrm{PWN}}/V_{\textrm{PWN}}$ \citep{2012MNRAS.427..415M}.

Maximum energy of CR electrons in  PWNe is determined by the balance of acceleration
and loss processes. Characteristic acceleration time of CR with energy $E$ and
Larmor radius $r_\textrm{L}=E/ZeB$ in magnetic field $B$ in diffusive shock
acceleration at SNR shock wave of velocity
$V_{\textrm{sw}}=\beta_{\textrm{sw}}c$  in Bohm diffusion regime  
with diffusion coefficient 
$D_\textrm{B}= r_\textrm{L} c/3$   is  
$t_{\textrm{acc}}=8 D_\textrm{B} / V^2_{\textrm{sw}} = 8 r_\textrm{L} /3\beta^2_{\textrm{sw}}c$ \citep{2021PhRvD.103h3010E}.
Similar value $t_{\textrm{acc}}\approx r_\textrm{L}/c$ is expected in PWNe
with relativistic TS \citep{2017SSRv..207..319P}. For most important 
synchrotron losses maximum energy of CR electron 
$E_{e,\textrm{max}}= \gamma_{e,\textrm{max}}m_ec^2$ follows from equality
$t_{\textrm{acc}}=t_{\textrm{syn}}$:
\begin{equation} \label{emax}
\gamma_{e,\textrm{max}} = \frac{3m_ec^2}{2e^{3/2}B^{1/2}} =3.8\times 10^{10}B^{-1/2}_{-5}. 
\end{equation}
Observable spectra of leptonic $\gamma$-ray emission are determined by 
parameters of leptonic  CR spectra (\ref{eq:23}) and background radiation
\citep{2018A&A...612A...2H}.

We note also that  in young SNRs, including a magnetar-conncted one, 
hadronic and leptonic emission are generated in a slow cooling regime.
Both the hadronic energy loss time for inelastic $pp$ collisions 
$t_{pp,\gamma} \approx 1.7\times 10^{5} (n_{\textrm{ISM}}/1 \textrm{ cm}^{-3})^{-1}
\textrm{ kyr}$ and  the leptonic cooling time  for synchrotron and IC channels 
$t_{\textrm{i}} 
\approx 1.3\times 10^{2}(w_{\textrm{rad,CMB}}/w_{\textrm{rad,i}})(E_{e}/10 \textrm{ TeV})^{-1}
\textrm{ kyr}$ (i = CMB, IR, SL) exceed the magnetar age $t_{\textrm{mag}}\approx 2$ kyr 
for $E_{e} \lesssim 100 \textrm{ TeV}$.

\subsection{Energy requirements for gamma-ray sources
in the SGR 1900+14 region}\label{s3-2}
 
To begin with, we assess  the energy requirements for potential 
sources of observed GeV-TeV emission from SGR 1900+14 outskirts. 

To this end,  we use the recipe proposed 
in \cite{2018A&A...612A...3H}. In the leptonic scenario only 
IC emission from CMB photon scattering in Thomson regime is considered.

In  the hadronic emission scenario,  the $\gamma$-ray emission from 
HOTS J1907+091 can be explained as a result of pp interactions of
SNR/SFR  shock-accelerated protons and heavy nuclei with  the target
nuclei of ISM plasma.

For the  mentioned above  HOTS J1907+091 integral photon  flux
$F_{\textrm{ph}}(\varepsilon > 1 \textrm{ TeV}) = 4.3 \times 10^{-13}
\textrm{ cm}^{-2} \textrm{ s}^{-1}$
the corresponding integral energy flux  is 
$F_{\varepsilon}(\varepsilon > 1 \textrm{ TeV}) = 2.3 \textrm{ TeV} \times F_{\textrm{ph}}
(\varepsilon > 1 \textrm{ TeV})$ $=9.9 \times 10^{-13} \textrm{ TeV} 
\textrm{ cm}^{-2}\textrm{ s}^{-1}$ for the spectral index $\Gamma = 2.3$.
The necessary energy of accelerated CR  protons with $E_{p}>10$ TeV  is:
\begin{equation} \label{eq:1}
    \begin{split}
        W_{\textrm{cr},p}(E_{p}>10 \textrm{ TeV}) \sim 4\pi d^2 F_{\varepsilon}(\varepsilon >1 \textrm{ TeV})t_{pp,\gamma} \\
        \sim  1.3\times 10^{50}d_{\textrm{10kpc}}^{2} (n_{\textrm{ISM}}/(1 \textrm{ cm}^{-3})^{-1}
        \textrm{ erg},   
    \end{split}
\end{equation}
or the total CR proton  energy 
\begin{equation} \label{eq:2}
\begin{split}
W_{\textrm{cr},p}=W_{\textrm{cr},p}(E_{p}>1 \textrm{ GeV})\\
\sim (1 \textrm{ GeV}/10 \textrm{ TeV})^{2-\Gamma_{\textrm{cr},p}} 
\times W_{\textrm{cr},p}(E_{p}>10 \textrm{ TeV})\\
\sim 1.9\times 10^{51} d_{\textrm{10kpc}}^{2}(n_{\textrm{ISM}}/1 \textrm{ cm}^{-3})^{-1} \textrm{ erg}
 \end{split}
\end{equation}
for 1 GeV $< E_{p} <$ 1 PeV CR bubble with power-law spectrum (\ref{eq:20}) 
and with  the spectral index $\Gamma_{\textrm{cr},p}$  equal to the spectral 
photon index $\Gamma$, as is expected in hadronic mechanism. We use  here 
the value $\Gamma=2.3$ that corresponds to the average index of known
Galactic VHE $\gamma$-ray sources \citep{2018A&A...612A...1H} and  
the  typical distance to the sources $d = 10$ kpc. 

Similar estimate of $W_{p,\textrm{tot}}$ for HOTS J1907+091 flux follows from the well-known formula for the standard chemical
composition of CRs and the ambient gas \citep{2004vhec.book.....A}
$F_{\textrm{ph}}(\varepsilon >1 \textrm{ TeV})
\approx 0.2\times 10^{-11}A \textrm{ cm}^{-2}\textrm{ s}^{-1}$ for
$\Gamma_{\textrm{cr},p} = 2.3$, where $A$ is the scaling parameter 
\begin{equation}
    A = \frac{W_{\textrm{cr}, p}}{10^{50} \textrm{ erg}} \left( \frac{d}{1 \textrm{ kpc}} \right) ^{-2}
    \frac{n_{\textrm{ISM}}}{1 \textrm{ cm}^{-3}}.
\end{equation}

In the leptonic emission scenario with IC scattering
of  CMB  radiation  by  SNR/SFR/PWN shock-accelerated electrons for the 
HOTS J1907+091 integral energy  flux the necessary energy of
accelerated leptons is expected to be
\begin{equation} \label{eq:4}
    \begin{split}
        W_{\textrm{cr},e} (E_{e}>10 \textrm{ TeV}) \sim 4\pi d^2 F_{\varepsilon}(\varepsilon >1 \textrm{ TeV})t_{\textrm{IC,CMB}} \\
        \sim W_{\textrm{cr},p}(E_{p}>10 \textrm{ TeV}) (t_{\textrm{IC,CMB}}/t_{pp,\gamma}) \\
        \sim 1.3\times 10^{47} d_{\textrm{10kpc}}^{2}\textrm{ erg},
    \end{split}
\end{equation}
or the  total CR lepton energy
\begin{equation} \label{eq:5}
    \begin{split}
W_{\textrm{cr},e}= W_{\textrm{cr},e} (E_{e}>1 \textrm{ GeV})\sim 16 W_{\textrm{cr},e} (E_{e}>10 \textrm{ TeV}) \\
\sim 1.9\times 10^{48} d_{\textrm{10kpc}}^{2}\textrm{ erg} .   
    \end{split}
\end{equation}
In the HE (100 MeV -- 100~GeV) band the {\it Fermi}-LAT extended source
4FGL J1908.6+0915e, overlapping the  SGR 1900+14 position, 
is modelled by a power-law spectrum with integral energy flux
$F_{\varepsilon}(\varepsilon > 100 \textrm{ MeV}) = 
2.75 \times 10^{-11} \textrm{ erg} \textrm{ cm}^{-2} \textrm{ s}^{-1}$ 
and photon index $\Gamma=2.2$ \citep{2020ApJS..247...33A}. In the 
hadronic emission scenario with photon
energy $\varepsilon \approx 0.1 E_{p}$ a requirement  for the total
CR proton energy 
\begin{equation} \label{eq:6}
    \begin{split}
    W_{\textrm{cr},p}=W_{\textrm{cr},p}(E_{p}>1 \textrm{ GeV})\\
    \sim 1.3\times 10^{51}d_{\textrm{10kpc}}^{2} (n_{\textrm{ISM}}/1 \textrm{ cm}^{-3})^{-1} \textrm{ erg}
     \end{split}
\end{equation}
is similar to the  considered above HOTS J1907+091 case owing to 
an agreement of HE and VHE spectra of SGR 1900+14 
(Fig.~\ref{fig:PionIC_PL}). 

In the leptonic-CMB emission scenario the 100 MeV -- 100~ GeV 
$\gamma$-ray band corresponds to 
200 GeV $\lesssim E_{e}  \lesssim $  6\,TeV  and a requirement  for the
CR $E_{e} > 200$ GeV electron energy is 
\begin{equation} \label{eq:7}
    \begin{split}
        W_{\textrm{cr},e} (E_{e} > 200 \textrm{ GeV}) \sim 4\pi d^2 F_{\varepsilon}(\varepsilon>100 \textrm{ MeV}) \\
        \times t_{\textrm{IC,CMB}}(E_{e}=200 \textrm{ GeV}) \sim 6.4\times 10^{49} d_{\textrm{10kpc}}^{2}\textrm{ erg},
    \end{split}
\end{equation}
and for the total CR electron energy we have
$W_{\textrm{cr},e}=W_{\textrm{cr},e} (E_{e}>1 \textrm{ GeV}) 
\sim 1.9  \times 10^{50}d_{\textrm{10kpc}}^{2}\textrm{ erg}$.

The obtained estimates of the required CR energy in the observed 
HE -- VHE sources, namely, $W_{\textrm{cr},p} \sim 1.3\times 10^{51}
d_{\textrm{10kpc}}^{2} (n_{\textrm{ISM}}/1 \textrm{ cm}^{-3})^{-1} \textrm{ erg}$
for hadrons and $W_{\textrm{cr},e}\sim 1.9\times 10^{50}d_{\textrm{10kpc}}^{2}
\textrm{ erg}$ for leptons  correspond  to a ordinary SNR/PWN case for
sources at distances of 2-5 kpc, but suggest  a HNR-type of energy 
reserve for distances over 10 kpc.

\subsection{Modelling the multiband spectral energy distribution (SED)
of the  SGR 1900+14 region in SNR case}\label{s3-3}

In SNR and SFR models of the multiband spectral energy distribution 
of the  SGR 1900+14 neighbourhood the observed $\gamma$-ray emission 
consists of hadronic 
contribution from proton (i = $p$) and leptonic contribution from 
electron (i = $e$)  CRs  with spectrum (\ref{eq:20})
and with its   parameters  
$E_{\textrm{cr,i,min}}$, $E_{\textrm{cr,i,max}}$, $N_{\textrm{0,i}}$,
$E_{\textrm{cr,i,0}}$, $E_{\textrm{cr,i,cut}}$, $\Gamma_{\textrm{cr,i,1}}$.
Additional parameters are electron-to-proton energy ratio  
$K_{ep}$ and external shell number density $n_{\textrm{sh}}$.

Shock accelerated electron CRs will contribute to the synchrotron 
radio to X-ray emission and also 
to the observed VHE $\gamma$-ray emission via IC scattering off 
background photons in both upstream and downstream regions.
Besides the CMB component with temperature
$T_{\textrm{CMB}}=2.7 \textrm{ K}$ and energy density 
$w_{\textrm{CMB}}=0.26 \textrm{ eV} \textrm{ cm}^{-3}$  we consider an infrared
(IR, $T_{\textrm{IR}}=107 \textrm{ K}$, 
$w_{\textrm{IR}}=1.19 \textrm{ eV} \textrm{ cm}^{-3}$) and a starlight
(SL, $T_{\textrm{SL}}=7906 \textrm{ K}$, 
$w_{\textrm{SL}}=1.92 \textrm{ eV} \textrm{ cm}^{-3}$) ones as the  representative
values for the population of TeV PWNe in the HGPS \citep{2018A&A...612A...2H}. 

\begin{figure*}
    \centering
    \includegraphics[width=14cm]{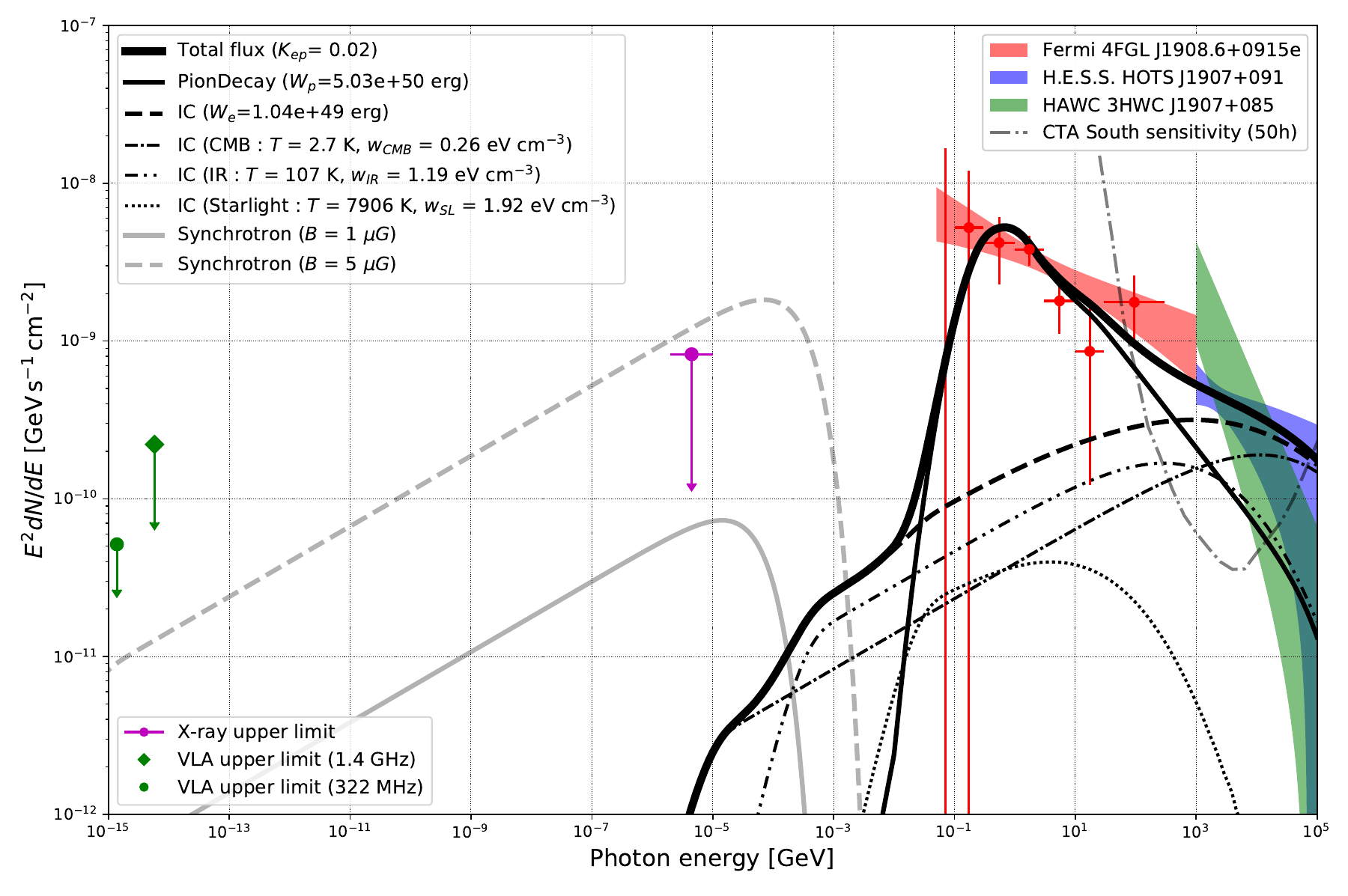}
    \caption{
    Modelled SED of the region of the magnetar SGR 1900+14 in SNR model.
    NAIMA-fitted parameters are given in Table 1 (the case of SNR at 
    distance $d=12.5$ kpc  and shock accelerated proton and electron 
    power-law     spectra with fixed $K_{ep}=0.02$). Observations of 
    \fermilat 4FGL J1908.6+0915e (in red),
    H.E.S.S. HOTS J1907+091 (in blue) and HAWC 3HWC J1907+085 
    (in green) are presented together 
    with upper limits in the  X-ray (in magenta) and  the radio
    (in green) bands. The total $\gamma$-ray
    flux is a sum of  hadronic ($pp$ collisions with subsequent 
    neutral pion decay) and leptonic 
    (IC scattering of electrons on CMB, IR and SL background photons) contributions. Synchrotron 
    emission is calculated for two values of the ambient magnetic field -- 1 $\mu$G and 5 $\mu$G. 
    The grey dash-dotted line represents the sensitivity of the CTA 
    South array for a zenith angle  $z=40^{\circ}$.}
    \label{fig:PionIC_PL}
\end{figure*}

\begin{figure*}
    \centering
    \includegraphics[width=14cm]{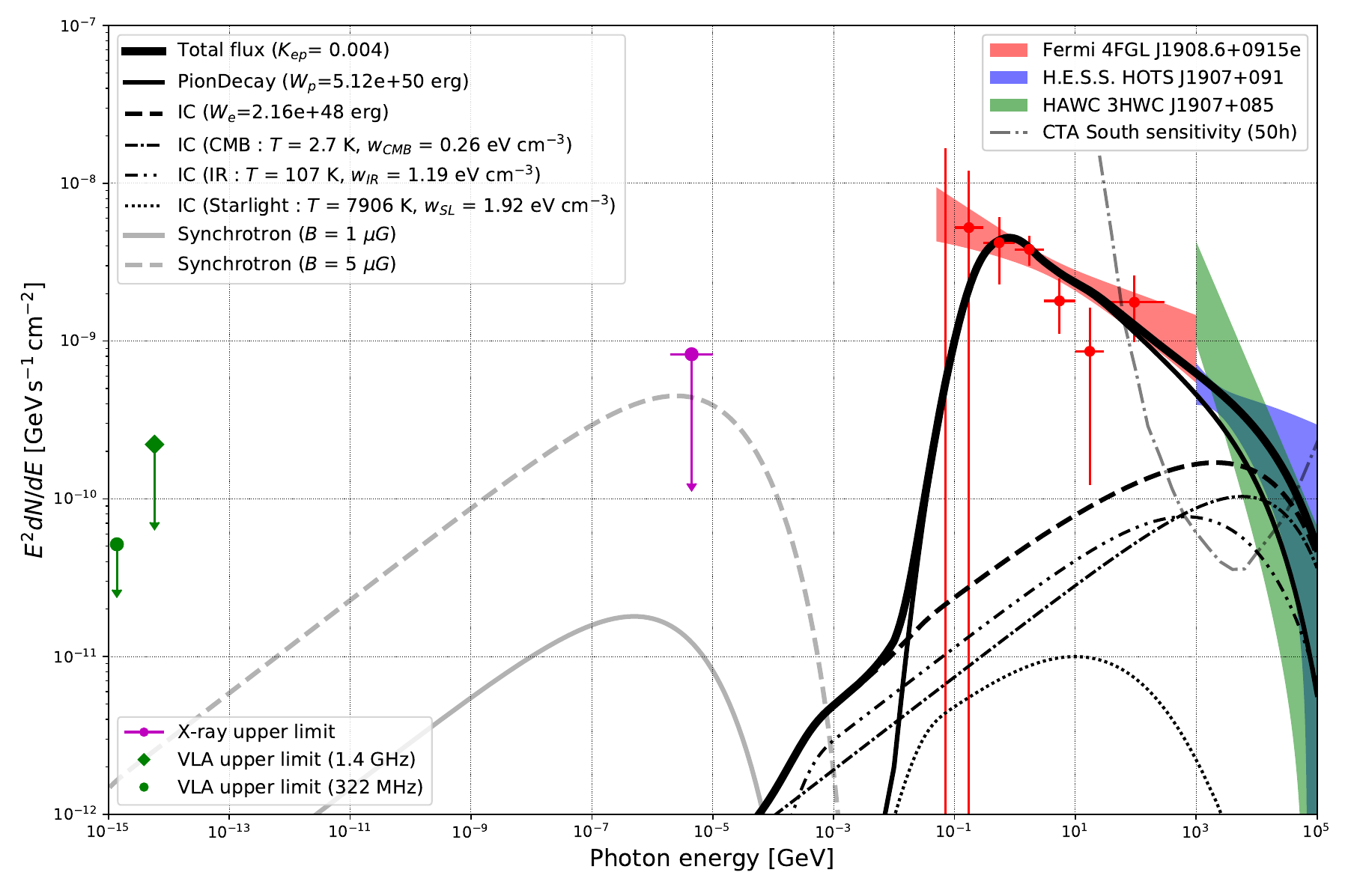}
    \caption{
    Same as in Fig.~\ref{fig:PionIC_PL} but for exponential cut-off power-law spectra with free $K_{ep}$.}
    \label{fig:PionIC_ECPL}
\end{figure*}

\begin{figure*}
    \centering
    \includegraphics[width=14cm]{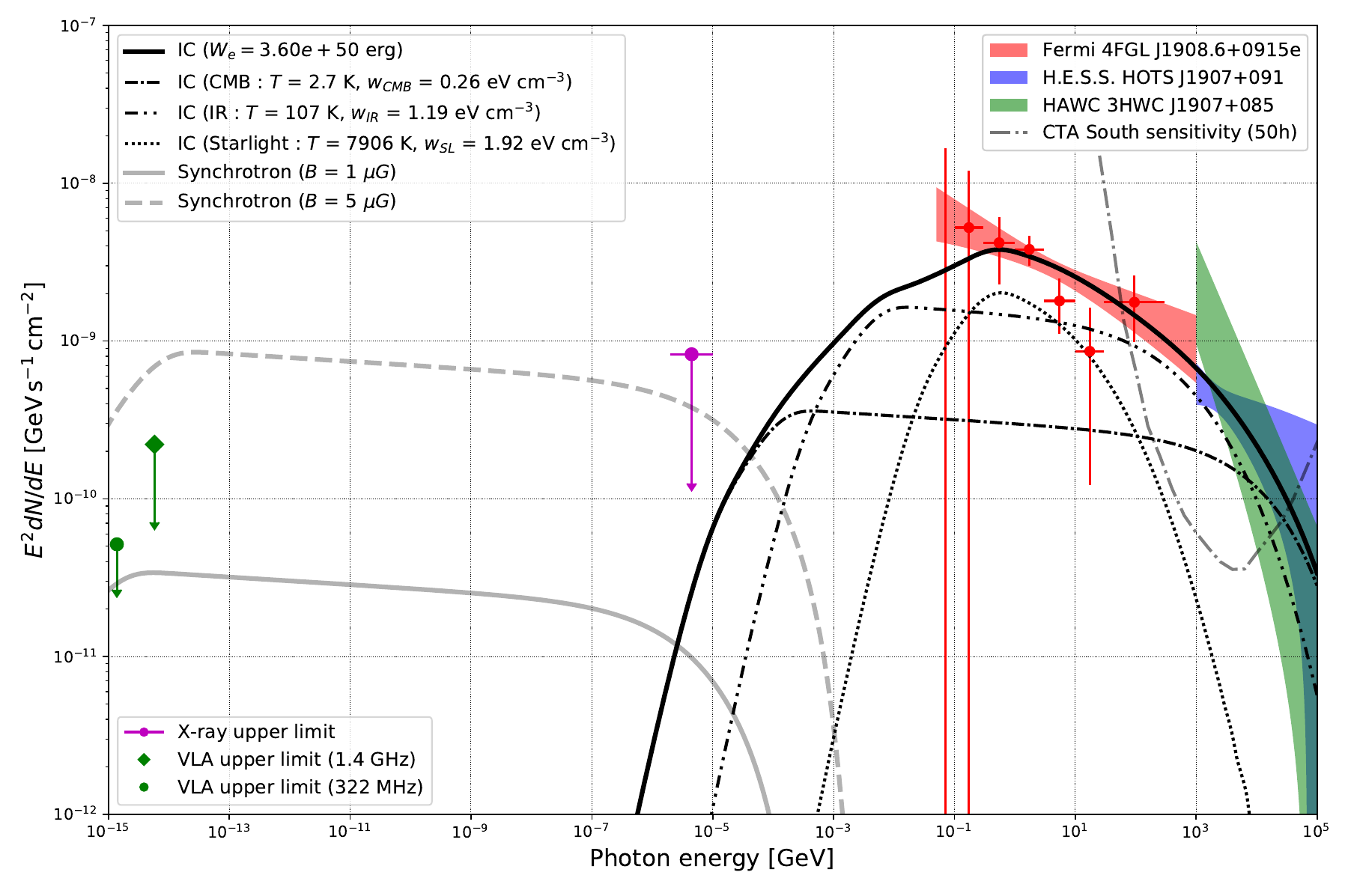}
    \caption{Modelled SED of the region of the magnetar SGR 1900+14 in PWN model.
    NAIMA-fitted parameters are given in Table 1 (the case of PWN at distance $d=12.5$ kpc and  termination shock/reconnection 
    accelerated leptons with an exponential cut-off broken power-law spectrum).  
    The total $\gamma$-ray flux is a sum of contributions from IC scattering of leptons on CMB, IR and SL background photons. The rest of the data is the same as in Fig.~\ref{fig:PionIC_PL}.
  }
    \label{fig:PWN_BPL_IC}
\end{figure*}

In order to explain observed by 
\fermilat (extended source 4FGL J1908.6+0915e)  and HESS 
(extended source HESS J1907+089/HOTS J1907+091) fluxes
\footnote{ As we noted earlier the  HAWC source 3HWC J1907+085 corresponds
to the H.E.S.S. source HESS J1907+089/HOTS J1907+091} 
we used the  NAIMA package for computation of non-thermal radiation from CR 
populations \citep{2015ICRC...34..922Z}.
NAIMA performs Markov Chain Monte Carlo (MCMC) fitting of 
radiative models to  observed
X-ray - TeV $\gamma$- ray spectra and  derives the best-fit 
and uncertainty distributions 
of spectral model parameters through MCMC sampling of their 
likelihood distributions. 
In our case \fermilat observations include only seven bins with an overall significance 
only 4.58 $\sigma$ and an analytical power-law approximation of differential flux density
of 4FGL J1908.6+0915e in the 50 MeV – 1 TeV energy range. H.E.S.S. observations of 
HESS J1907+089/HOTS J1907+091 are of even worse quality – only the measured integral 
photon flux $F_{\textrm{ph}}(\varepsilon > 1 \textrm{ TeV}$  is presented, 
which we have analytically approximated by power-law 
differential flux density in 1 – 10 TeV (see Section \ref{s2-1} and
Fig.~\ref{fig:PionIC_PL}-\ref{fig:PWN_TwoPopulations_IC} for details).
Therefore as input data for NAIMA fitting we have taken fluxes and their  uncertainties
for twenty bins of \fermilat analytical approximation in the 50 MeV – 1 TeV range  and   
for twenty bins of H.E.S.S.  approximation in the 1 TeV – 10 TeV  range. Fitted CR spectra
was taken of a  power-law (PL) and  an exponential cut-off power-law   (ECPL) 
type (\ref{eq:20}) with fixed values of parameters $E_{\textrm{cr,i,min}}$, $E_{\textrm{cr,i,max}}$ 
and $K_{ep}$ (only for PL spectrum) (Table~\ref{parameters}).

Estimated best fit CR parameters and wind shell nucleon density are presented in
(Table~\ref{parameters}) and  corresponding best fit spectral energy
distribution - in Fig.~\ref{fig:PionIC_PL}-\ref{fig:PionIC_ECPL}.
As it was expected from analytical estimates, the main contribution
in the observed $\gamma$-ray 
flux corresponds to the hadronic mechanism with  a typical for ordinary 
relatively close ($d\lesssim 5$ kpc)  SNR value of CR energy reserve, 
but with  high --   of Hypernova type  -- value of CR energy for 
distant (($d > 10 $ kpc) SNR
$W_{\textrm{cr},p}\approx 3\times 10^{50}d_{\textrm{10kpc}}^{2}(n_{\textrm{sh}}/10 
\textrm{ cm}^{-3})^{-1}$ erg even 
for swept-up shell with enhanced density  $(n_{\textrm{sh}}
\sim 10 \textrm{ cm}^{-3})$. 
This result is weakly sensitive to the electron contribution
($K_{ep}$), because the 
main energy losses occur in the 1 -- 10 GeV  range. On the other hand, 
the contribution of electron CR 
may be important and even dominant in TeV band for softer
(increasing $\Gamma_{\textrm{cr,i}}$) spectra.
As it follows from (\ref{fkn}), KN depression in background 
radiation of temperature $T$  starts 
at electron Lorentz factor $\gamma_{e,\textrm{KN}}$ such that
$f_{\textrm{KN}}\lesssim 0.5$ or at IC photon energy
$\varepsilon_{\textrm{IC,KN}} \approx 4 k_BT\gamma_{e,\textrm{KN}}
\approx 73.3 \times 10^{13}(T/T_{\textrm{CMB}})^{-1}$ eV. Therefore
in our case contribution of IC scattering off SL photons is negligible
in TeV band.

Harder spectra ($\Gamma_{\textrm{cr,i}} \sim 2.2 - 2.4$) can 
explain observations without electron CR
contribution, but at the cost of increasing the total CR energy. 
It is important to note that
hadronic flux  is proportional to the product
$W_{\textrm{cr},p}n_{\textrm{sh}}$, i.e., we can decrease necessary 
total  proton CR energy increasing shell density   $n_{\textrm{sh}}$, 
but observations (non-detection of molecular cloud) put restriction
$n_{\textrm{sh}}\lesssim 10^2$ cm$^{-3}$. 

Synchrotron radiation of electron CRs in radio to X-ray band is
determined by the value of 
magnetic field in region, filled by CRs. Excluding a narrow 
acceleration region around shock front, 
magnetic field in the rarefied wind bubble and inside HNR is
expected to be  of order of  or less than 
typical interstellar value $B\lesssim 3 \mu G$.  Corresponding 
flux does not   exceed existing 
observational limits (Fig.~\ref{fig:PionIC_PL}-\ref{fig:PWN_TwoPopulations_IC}).

\subsection{Modelling the multiband SED of the SGR 1900+14 region in PWN  case}\label{s3-4}

\begin{figure*}
    \centering
    \includegraphics[width=14cm]{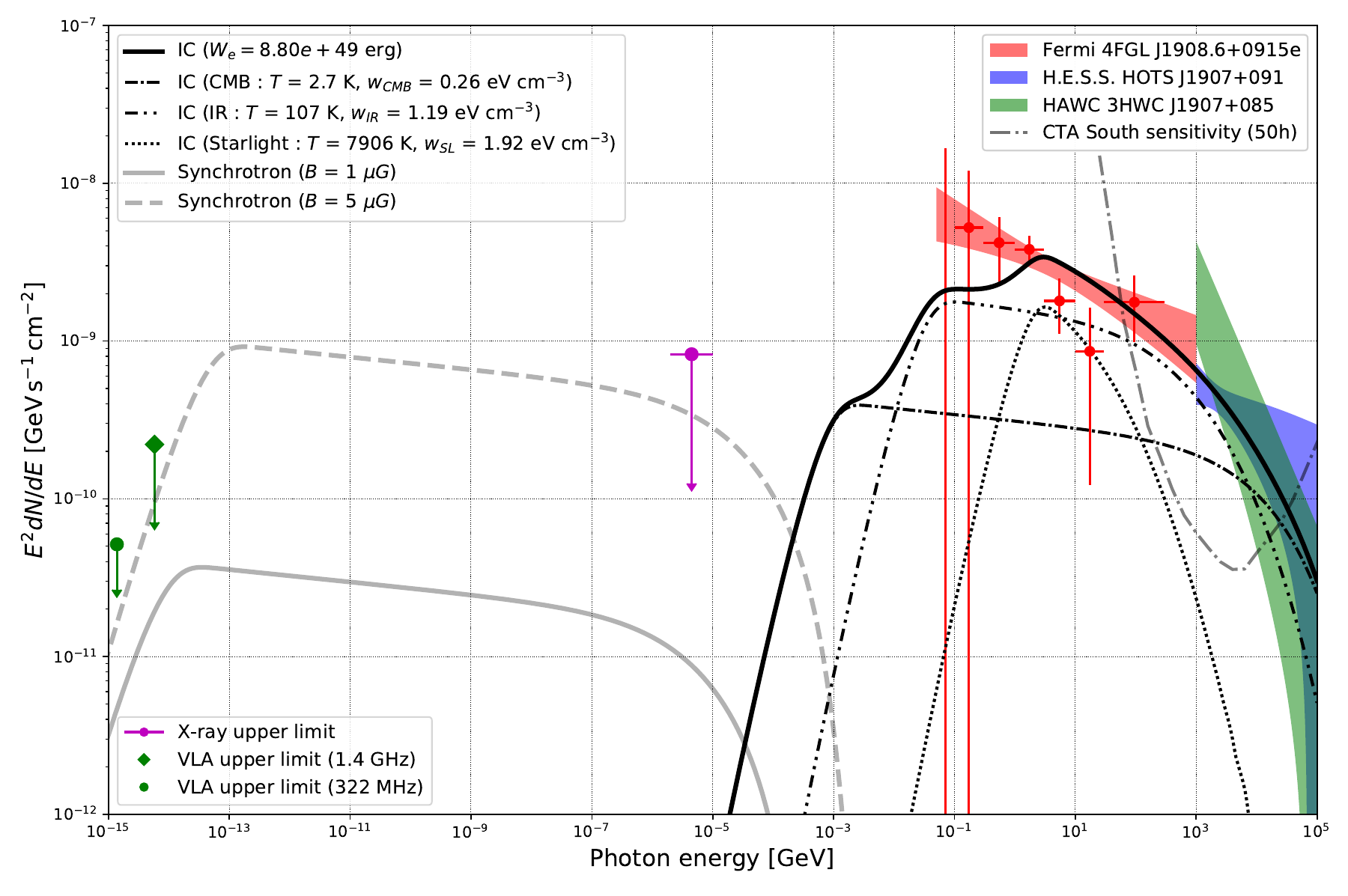}
    \caption{
    Same as in Fig.~\ref{fig:PWN_BPL_IC} but for alternative exponential cut-off power-law spectrum with $E_{\textrm{min}}=10$ GeV.  
    }
    \label{fig:PWN_ECPL_IC}
\end{figure*}

The results of the NAIMA-fitting of SED  of the SGR
1900+14 neighbourhood in the PWN model for ECBPL  spectrum is presented
in Fig.~\ref{fig:PWN_ECPL_IC}, the estimated necessary leptonic  
CR parameters are presented in Table~\ref{parameters}. As we can see, 
similarly to the SNR case, necessary leptonic CR energy reserve 
$W_{\textrm{cr},e}\approx 2.3\times 10^{50} d_{\textrm{10kpc}}^{2}$ erg  
is typical for ordinary relatively close ($d\lesssim 5$ kpc)  PWNe, but increases by an order of magnitude  --  up to Hypernova-related  MWN type  -- for distant ($d > 10 $ kpc) PWNe.

For a decreasing  $\varepsilon^2F_{\textrm{ph}}(\varepsilon)$ flux 
in sub-TeV -- TeV region (a photon index $\Gamma >2$) 
an effective (average) spectral
index of   CR electrons should be $\Gamma_{\textrm{cr},e,\textrm{2}}
= 2\Gamma -1 > 3$. In our case we observe only this post-maximum decreasing 
$\varepsilon^2F_{\textrm{ph}}(\varepsilon)$ flux in the all $\gamma$-ray 
GeV -- TeV region with  $\Gamma \sim $ 2.2  (in the GeV band) -- 2.9 
(in the TeV band),
therefore we use for modelling the PWN broad band spectrum also 
an inspired by "alternative model" of 
\cite{2017ApJ...838..142I} ECPL electron spectrum with 
$E_{\textrm{min}} = E_{\textrm{br},e} \sim 10 \textrm{ GeV}$ 
and $\gamma_{\textrm{2},e} >$ 3 in order to estimate the minimum 
energy of leptonic CR, neceassary to 
explain observations. The results of the NAIMA-fitting of SED  
for alternative ECPL lepton spectra are presented  in
Fig.~\ref{fig:PWN_ECPL_IC} and necessary leptonic  
CR parameters are presented in Table~\ref{parameters}. Minimum 
leptonic CR energy reserve  
$\sim 6\times 10^{49}d_{\textrm{10kpc}}^{2}$ erg
is more than an order of magnitude lower than predicted in
millisecond magnetar-related MWN model.

\begin{figure*}
    \centering
    \includegraphics[width=14cm]{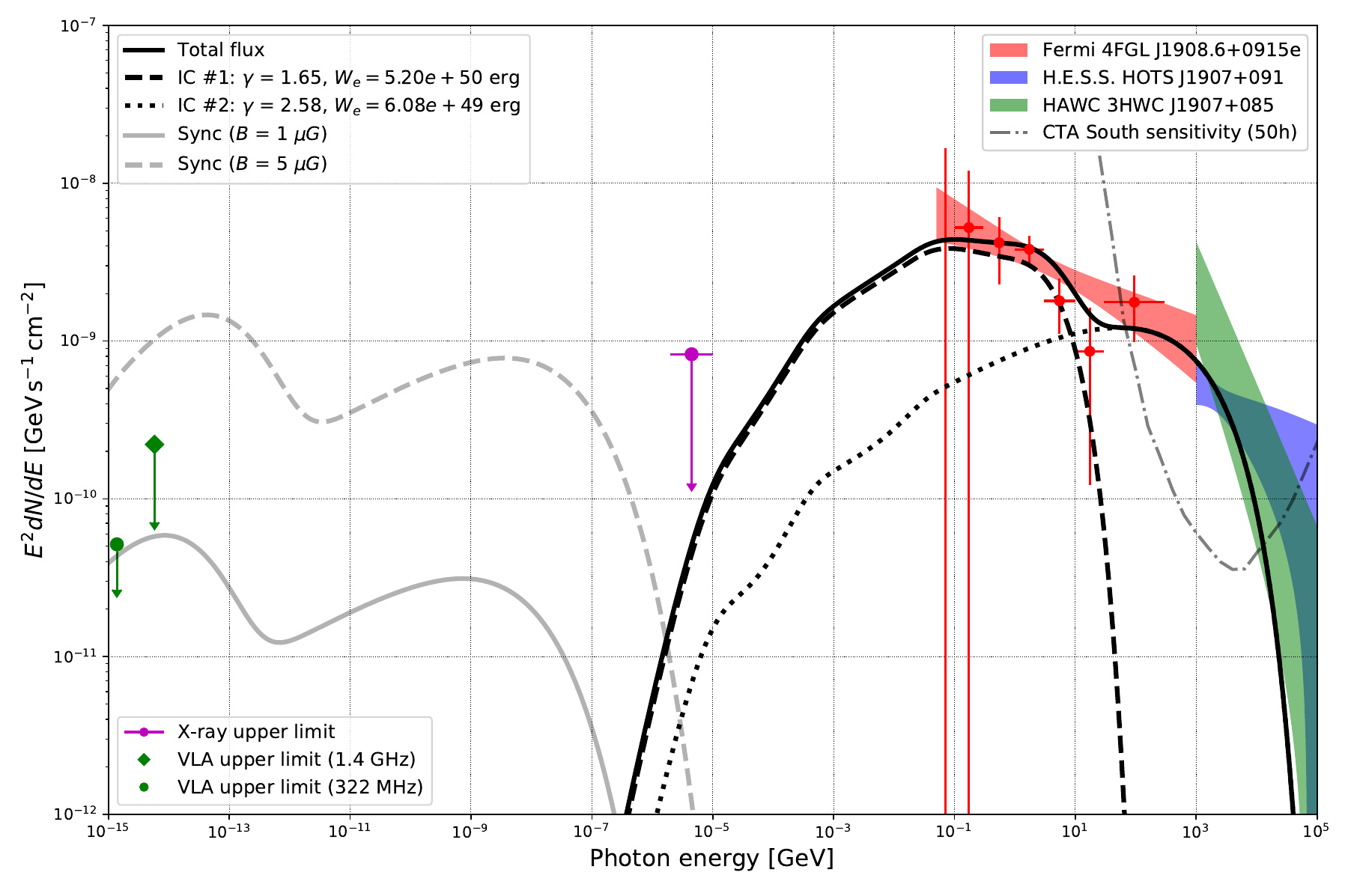}
    \caption{
    Same as in Fig.~\ref{fig:PWN_BPL_IC} but for two electron populations 
    with exponential cut-off power-law spectra. For both populations only 
    the total $\gamma$-ray fluxes as  sums  of contributions from IC scattering 
    of leptons on CMB, IR and SL background photons are presented.
    }
    \label{fig:PWN_TwoPopulations_IC}
\end{figure*}

To examine  a two lepton population model we consider also the case
of two (i = 1, 2) lepton populations with power-law exponential cut-off spectra 
\begin{equation} \label{eq:24}
N_{\textrm{i}}(E) = N_{\textrm{0,i}}(E/E_{\textrm{0,i}})^{-\gamma_{i,e}}
\exp(-E/E_{\textrm{cut,i}}),
\end{equation}
with $\gamma_{\textrm{1},e}< 2$ typical for PWN lepton CR spectra  and 
$\gamma_{\textrm{2},e} > 2$ typical for SNR electron CR spectra. 
SED for this case are presented in Fig.~\ref{fig:PWN_TwoPopulations_IC}
and leptonic  CR parameters -- in Table~\ref{parameters}. As it follows from 
Table~\ref{parameters}, the explanation of TeV $\gamma$-ray emission needs 
six time larger electron CR energy than can be provided by external shock 
accelerated electron CRs with $K_{ep}=0.02$ and cannot dominate in our case. 
But it still can be detected by CTA-like detectors. 

\begin{table*}
\caption{SNR and PWN NAIMA-fitted models of SED from the region of the magnetar SGR 1900+14 
\label{parameters}}
\resizebox{\textwidth}{!}{%
\begin{tabular}{lcccccc}
\hline
\multicolumn{1}{c}{\multirow{2}{*}{Parameter}} & \multicolumn{2}{c}{SNR: PL and ECPL spectra (i = $p$)} & \multicolumn{2}{c}{PWN: ECBPL and alternative ECPL spectra (i = $e$)} & \multicolumn{2}{c}{PWN:Two electron population spectra (i = $e$)} \\ \cline{2-7} 
\multicolumn{1}{c}{}                           & PL                 & ECPL               & ECBPL                           & ECPL                    & ECPL \#1                & ECPL \#2                \\ 
\hline
$E_{\textrm{cr,i,min}}$ {[}GeV{]}(fixed)                              & 1                       & 1                      & 1                             & 10                            & 1                            & 1                   \\ \hline
$E_{\textrm{cr,i,max}}$ {[}GeV{]}(fixed)                              & 1e6                     & 1e6                    & 1e6                           & 1e6                           & 1e6                          & 1e6              \\ \hline
$N_{\textrm{0,i}}$ {[}1/ eV{]}                                & (1.18$\pm$0.04)e35      & (1.42$\pm$0.05)e37     & (1.90$\pm$0.12)e40            & (8.91$\pm$0.50)e37            & (4.41$\pm$0.51)e38           & (8.81$\pm$0.44)e34     \\ \hline
$E_{\textrm{cr,i,0}}$ {[}TeV{]}                              & 3.93$\pm$0.1            & 0.78$\pm$0.01          & 0.19$\pm$0.01                 & 0.175$\pm$0.003               & 1.91$\pm$0.19                & 1.81$\pm$0.19             \\ \hline
$E_{\textrm{cr,i,br}}$ {[}TeV{]}                          & -                       & -                      & 0.0047$\pm$0.0003             & -                             & -                            & -                             \\ \hline
$E_{\textrm{cr,i,cut}}$ {[}TeV{]}                            & -                       & 185.2$\pm$9.5          & 396.7$\pm$41.7                & 424.3$\pm$21.1                & 0.0096$\pm$0.0008            & 9.99$\pm$1.06            \\ \hline
$\Gamma_{\textrm{cr,i,1}}$                                   & 2.55$\pm$0.01           & 2.41$\pm$0.03          & 1.49$\pm$0.07                 & -                             & 1.65$\pm$0.11                & -                       \\ \hline
$\Gamma_{\textrm{cr,i,2}}$                                   & -                       & -                      & 3.04$\pm$0.06                 & 3.08$\pm$0.03                 & -                            & 2.58$\pm$0.17            \\ \hline
$W_{p}$ {[}erg{]} (calculated$^{\dagger}$)            & 5.03e50                  & 5.12e50                 & -                  & -                             & -                             & -                    \\ \hline
$W_{e}$ {[}erg{]} (calculated$^{\dagger}$)             & 1.04e49                  & 2.16e48                 & 3.60e50           & 8.80e49                             & 5.20e50                       & 6.08e49              \\ \hline
$n_{\textrm{sh}}$ {[}cm$^{-3}${]}                            & 11.42$\pm$0.66          & 9.81$\pm$0.35          & -                             & -                             & -                            & -                   \\ \hline
$K_{ep}$                                         & 0.02 (fixed)            & 0.0041$\pm$0.0002      & -                             & -                             & -                            & -                      \\ \hline
$\chi^2/ndf$$^{\star}$                                  &42.10/36                          &30.74/34                       &16.54/34                             &40.24/36                                & \multicolumn{2}{c}{36.64/32}                     \\ \hline
\multicolumn{6}{l}{$^{\star}$ Best fit of 20 bins of \fermilat PL 0.05-1000 GeV spectrum and 20 bins of HOTS PL 1-10 TeV spectrum approximation}\\
\multicolumn{6}{l}{$^{\dagger}$ for distance to the source $d=12.5$ kpc}
\end{tabular}%
}
\end{table*}

\section{Possible sources of HE -- VHE gamma-ray emission from SGR 1900+14 region}\label{s4}

\subsection{Supernova Remnants}\label{s4-1} 

As follows from the above  mentioned data, magnetar SGR 1900+14 was born about 
2 kyr ago as a result of $M\approx 17 M_{\odot}$ progenitor SN outburst
inside the $M\sim 10^3-10^4 M_{\odot}$ young star cluster  Cl 1900+14/SGR 1900+14. 

An ordinary SNR with an explosion energy $E_{\textrm{SN}}\sim 10^{51}$ erg
and ejected mass $M_{\textrm{ej}}\sim 5 M_{\odot}$
in a dense ($n_{\textrm{mc}}=\rho_{\textrm{mc}}/\mu \sim 10^4 \textrm{ cm} ^{-3}$) 
molecular cloud (clump) 
with the mean molecular weight $\mu=1.4m_{\textrm {H}}$ and of  radius of 
$R_{\textrm{mc}}\sim 0.5 \textrm{ pc}$ is expected to be at present at an  adiabatic
(Sedov - Taylor, ST) stage of evolution. The radius $R^*_{\textrm{ST}}$ and age $t^*_{\textrm{ST}}$ 
of SNR at  beginning of the ST stage is determined by characteristic scales of length $R_{\textrm{ch}}$:
\begin{equation} \label{eq:16}
\begin{split}
    R^*_{\textrm{ST}}=c_1(n,s)R_{\textrm{ch}}=c_1(n,s)M_{\textrm{ej}}^{1/3}\rho_{\textrm{mc}}^{-1/3}=\\ 
    3.1c_1(n,s)(M_{\textrm{ej}}/M_{\odot})^{1/3}n_{\textrm{mc}}^{-1/3}\, \textrm{pc},
     \end{split}
\end{equation}
and time $t_{\textrm{ch}}$:
\begin{equation} \label{eq:17}
\begin{split}
    t^*_{\textrm{ST}}=c_2(n,s)t_{\textrm{ch}}=
    c_2(n,s)E_{\textrm{SN}}^{-1/2}M_{\textrm{ej}}^{5/6}\rho_{\textrm{mc}}^{-1/3}=\\
    4.2\times 10^2c_2(n,s)E_{\textrm{SN,51}}^{-1/2}(M_{\textrm{ej}}/M_{\odot})^{5/6}n_{\textrm{mc}}^{-1/3}\,
    \textrm{yr},
     \end{split}
\end{equation}
where $c_1(n,s)\sim 1$ and $c_2(n,s)\sim 1$ are dimensionless functions of  power-law 
density profiles of an ejecta with a power-law index $n$ and an ambient medium with 
a power-law index $s$ \citep{1999ApJS..120..299T}.

At the ST stage the time-depended external (forward) shock wave radius $R_{\textrm{ST}}$ and 
the velocity $D_{\textrm{ST}}$ are given by   
\begin{equation} \label{eq:8}
    \begin{split}
        R_{\textrm{ST}}(t)= (2.026E_{\textrm{SN}}/\rho_{\textrm{mc}})^{1/5}t^{2/5} \\
        = 0.791 (E_{\textrm{SN,51}}/n_{\textrm{mc,4}})^{1/5}
        t_{\textrm{ kyr}}^{2/5} \textrm{ pc},
    \end{split}
\end{equation}

\begin{equation} \label{eq:9}
    \begin{split}
        D_{\textrm{ST}}(t)=(2/5)R_{\textrm{ST}}(t)/t
        = (2/5)(2.026E_{\textrm{SN}}/\rho_{\textrm{mc}})^{1/2}R_{\textrm{ST}}(t)^{-3/2} \\
        = 2.15\times 10^7 (E_{\textrm{SN,51}}/n_{\textrm{mc,4}})^{1/2}R_{\textrm{ST,pc}}(t)^{-3/2} 
        \textrm{cm} \textrm{ s}^{-1} \\
        = 3.07\times 10^7 (E_{\textrm{SN,51}}/n_{\textrm{mc,4}})^{1/5}
        t_{\textrm{ kyr}}^{-3/5} \textrm{ cm} \textrm{ s}^{-1}
    \end{split}
\end{equation}
\citep{1999ApJS..120..299T, 2008ARA&A..46...89R}.

But observations do not confirm the presence of SNR inside molecular cloud.
Moreover,  even the molecular cloud itself is not observed. Recent modelling of
star formation shows that in a young compact cluster of massive
stars alike Cl 1900+14 at the moment of the first SN explosion the gas
component is practically completely
expelled from  star cluster by stellar winds of cluster's stars and is 
concentrated in wind shell at $\sim $ 40 pc.
Magnetar-connected SN exploded in rarefied mix of stellar winds   
and, additionally energised by newborn millisecond magnetar,  has reached a
large size by now (see discussion in Section \ref{s4}).  

In our Hypernova model the HNR with $E_{\textrm{HNR}} \sim 10^{52} \textrm{ erg}$
evolves inside the low-density wind bubble 
($n_{\textrm{w}} \sim 10^{-3} \textrm{ cm}^{-3}$, $B_{\textrm{w}}
\lesssim 3 \,\mu \textrm{G}$) 
and now ($t_{\textrm{HNR}} \approx 2 \textrm{ kyr}$)
is at the ST stage with the radius $R_{\textrm{HNR}} \approx 35 \textrm{ pc}$ 
and the shock velocity $V_{\textrm{HNR}}=0.4R_{\textrm{HNR}}/t_{\textrm{HNR}}
\approx 7\times 10^{3} \textrm{ km} \textrm{ s}^{-1}$, 
and  is approaching an extended ($R\sim 40 \textrm{ pc}$) shell-like halo with
$n_{\textrm{sh}}\sim 10 \textrm{ cm}^{-3}$ and the total swept up mass
of  $\sim  10^4 {\textrm{M}}_{\odot}$ (see Section \ref{s4-2} for details).  

Analogously to discussed earlier SNR case,  the HNR shock is also effective
both  proton and electron CR
accelerator.  For total HNR shock energy $W_{\textrm{HNR}}=10^{52}$ erg we expect
$W_{\textrm{cr},p}\sim 0.1\eta_{p,-1}W_{\textrm{SNR}} \sim 10^{51}\eta_{p,-1}
\textrm{ erg}$ in proton CRs and 
$W_{\textrm{cr},e} \sim 0.01K_{ep,-2} W_{\textrm{cr},p}\sim 10^{49}K_{ep,-2}\textrm{ erg}$ 
in electron CRs.

Maximum energy of accelerated CRs with the charge $Ze$ in the diffusive 
shock acceleration at
the HNR shock with post-shock magnetic field $B_{\textrm{HNR}}=\sqrt{11}B_{\textrm{w}}$
according to (\ref{eq:19}) is $E_{\textrm{cr,max}}\approx 9Z$ PeV. For proton CRs 
energy losses are small during HNR age, but for electron CRs radiate cooling is 
important for $E_{\textrm{cr},e}\geq 100$ TeV.

Owing to a low density of target particles inside the HNR,  we expect the main 
contribution to the observed VHE $\gamma$-ray emission of the SGR 1900+14 region 
from inelastic collisions of shock accelerated protons and heavier nuclei with target
protons and heavier nuclei of  dense shell-like boundary of stellar wind bubble halo with 
a subsequent neutral pion decay. Since the shock wave is close to the bubble
boundary, but still inside the bubble $R_{\textrm{HNR}}\lesssim R_{\textrm{swb}}$, only CR 
from upstream shock region  will effectively collide with target shell material.
Their energy reserve $W_{\textrm{cr},p,\textrm{up}}$  is of order of total CR energy
$W_{\textrm{cr},p,\textrm{up}}\sim W_{\textrm{cr},p}$ therefore in following 
we ignore this difference.

In considered  HNR model a natural explanation for large size of  young $\sim 2$ kyr
GeV-TeV source can be given. Present radius of HNR shock
$R_{\textrm{HNR}} \approx 35$ pc just corresponds to the size of the H.E.S.S. source
HOTS J1907+091 of the angular radius 
$\Theta_{\textrm{VHE}}=0^{\circ}.17\pm 0^{\circ}.04$,
that corresponds to the linear radius of 
$R_{\textrm{VHE}}\approx \Theta_{\textrm{VHE}}d_{\textrm{mag}} \approx 37\pm 8.7$ pc.
Hadronic component of TeV $\gamma$-ray emission is produced in the  wind bubble
shell of the enhanced number
density $n_{\textrm{sh}}$  and the  radius $R_{\textrm{sh}}\approx 40$ pc.
Accelerated at  the shock front  
electron and proton CRs diffusively expand ahead  the shock front 
at a characteristic distance \citep{2004vhec.book.....A}:
\begin{equation} \label{eq:22}
    L_{\textrm{dif}}\approx 2(D\times t)^{1/2} 
    \approx 35\left(\frac{E_{\textrm{cr}}}{10Z\,\textrm{GeV}}\times 
    \frac{3\mu G}{B}\right)^{0.24}t_{\textrm{kyr}}^{1/2} \textrm{  pc},
\end{equation}
where the best fitted diffusion coefficient $D$ is taken from the Galactic 
diffusion-reacceleration-convection model \citep{2017PhRvD..95h3007Y} and
complemented by dependence on magnetic field  in source $B$ \citep{1990acr..book.....B}
\begin{equation} \label{eq:21}
    D = D_0\left(\frac{\cal{R}}{4 \textrm{ GV}}\cdot
    \frac{3\mu G}{B}\right)^{\delta},
\end{equation}
with $D_0=6.14\times 10^{28} \textrm{ cm}^2 \textrm{ s}^{-1}$, $\delta=0.48$, and 
with the  slope of the injection spectrum  $\Gamma_{p,e}=2.37$ for CR rigidity
${\cal{R}} = E_{\textrm{cr}}/Ze > 16 \textrm{ GV}$ ($E_{\textrm{cr}}> 16Z$ GeV).

Eq.(\ref{eq:22}) implies that diffusion of  electron and proton CRs in upstream
region is efficient and proton CRs can penetrate in wind bubble shell and produce 
hadronic $\gamma$-ray emission even in case of tens-fold  enhanced value of magnetic
field $B$ in shock outskirts.

Explanation of a larger size  of  extended \fermilat HE $\gamma$-ray  source
4FGL J1908.6+0915e modelled by disc with the angular radius of
$\Theta_{\textrm{ HE}}=0^{\circ}.6$ or 
linear radius of $R_{\textrm{HE}}\approx
\Theta_{\textrm{ HE}}d_{\textrm{mag}} \approx 130$ pc is more problematic. 
Only proton and electron CRs with energies $E_{\textrm{cr}}\gtrsim 500$ GeV 
can reach such a distance in $t_{\textrm{HNR}}\approx 2$ kyr.  Taking into 
account mentioned above low overall detection 
significance of this extended source  (4.58$\sigma$), its large size 
may be a result of superposition of $\Theta_{\textrm{ VHE}}=0^{\circ}.17$ source 
around the  magnetar  and 
mentioned above source  4FGL J1910.2+0904c around star forming region  W49A.

\subsection{Star formation regions} \label{s4-2}

As mentioned above, two SFRs -- Cl 1900+14 and W49A -- can be potential 
contributors to the observed $\gamma$-ray emission from SGR 1900+14 region. 
But quantitative analysis of expected $\gamma$-ray fluxes from these sources
suggests the negligible level of their
contributions. In young magnetar SGR 1900+14-connected  star cluster 
Cl 1900+14 with only two luminosity-dominated  M5 Red Supergiant (RSG) 
stars  with masses $M\sim 14-18 M_{\odot}$ and  with  most likely the 
only SN outburst of progenitor of similar mass  (see discussion in 
Section \ref{s2-3}) we do not expect the presence of 
powerful CR accelerated shocks besides SNR one. SFR W49A, contrary
to the Cl 1900+14 case, 
is the most luminous  Galactic SFR at similar distance of $\sim  11$   kpc in
one of the most massive giant molecular cloud complex W49 with 
$M_{\textrm{gas}}\sim 10^6$  
$M_{\odot}$ \citep[][and references therein]{2013ApJ...779..121G}.
But H.E.S.S observations of W49A detected $\gamma$-ray emission only with  
the primary analysis, whereas the  additional 
cross-check analysis does  not confirmed detection level  (above 5 $\sigma$) 
\citep{2018A&A...612A...5H}. Detection of the $\gamma$-ray emission around W49A,
based on  \fermilat Pass 8 data  is claimed in \cite{Xin...Guo}. 

\subsection {Pulsar wind nebulae} \label{s4-3}

The are no active pulsars in the  SGR1900+14 region  and only a magentar-connected  relic MWN is a putative source.

Above we considered the $\gamma$-ray emission of HNR, energised by the spin-down
luminosity $L_{\textrm{sd}}\sim 10^{48} $ erg s$^{-1}$ of the newborn millisecond magnetar
with the initial spin-down  time-scale $t_{\textrm{sd}}\sim 10^4$ s. Total released energy 
$E_{\textrm{sd}}\sim 10^{52}$ erg  in the form of the  non-collimated  relativistic magnetar wind
is transformed in energy of electron-positron pairs,  accelerated at the termination 
shock and collected in MWN inside the expanding SN
ejecta. Adiabatic expansion of MWN results in an additional acceleration of SN ejecta 
to a Hypernova signatures at the expense of MWN cooling 
$E_{\textrm{MWN}}\propto R_{\textrm{HNR}}^{-1}$ for several weeks. 

But if the magnetar wind is well-collimated and jet-like
\citep{2007MNRAS.380.1541B,2021NewAR..9201614C}, it can drill the expanding ejecta 
and produce a long gamma-ray burst (LGRB) and a Type I superluminous SN 
(SLSN I)\citep{2003ApJ...589..871A,2009PhRvD..79j3001M,2019ApJ...871L..25P}. 
In this case the main part of magnetar rotational energy will be deposited in the
relativistic  Poynting-flux dominated  jet with 
the initial Lorentz factor $\Gamma_{\textrm{j,ini}}(r)>>1$, the initial
magnetization parameter 
$\sigma_{\textrm{j,ini}} (r) = B^2/4\pi\rho c^2>>1$ and the isotropic luminosity 
\begin{equation} \label{liso}
L^{\textrm{iso}}_\textrm{j} \approx L_{\textrm{sd}}\omega^{-1} 
\approx 4\pi r^2 \rho c^3 \gamma^2_{\textrm{j,ini}}(1+\sigma_{\textrm{j,ini}}),
\end{equation}
where  $\rho  = n_lm_e$ is the lepton rest  mass density
(all $B$, $\sigma_\textrm{j}$  and $\rho$ 
in the comoving frame), $\omega=\theta_\textrm{j}^2/4$ is the  isotropic 
correction, $\theta_\textrm{j}$ 
is the jet half-opening angle, $t_\textrm{j}=t_{\textrm{sd}}$ and 
$\Delta =ct_\textrm{j}$ are the duration and  the length of jet,
correspondingly. Interaction of 
the jet with ISM is analogous to the case of the GRB jet \citep{1999PhR...314..575P},
only in our case the matter of the jet consists dominantly of leptons, whereas
matter of the jet in a classical fireball model consists of  baryons (protons) and electrons. 
According to the fireball model, a jet with isotropic energy 
$E^{\textrm{iso}}=L^{\textrm{iso}}_\textrm{j}
t_\textrm{j}=\omega^{-1}E_{\textrm{rot}}$ interact with ISM  and 
creates a jet-like analog of MWN. It includes undisturbed ISM 
(region {\bf 1}) with the baryon (proton) number density
$n_{\bf 1}$, shocked by relativistic external  shock region {\bf 2} 
with the proton number density $n_{\bf 2}$,
shocked by relativistic reverse shock region {\bf 3} (MWN)  with the lepton number density $n_{l,\bf 3}$ 
(the equivalent proton number density $n_{\bf 3}= (m_e/m_p)n_{l,\bf 3}$
and the undisturbed jet region {\bf 4} ( the free magnetar wind) with 
the lepton number density $n_{l,\bf 4}$   (the equivalent proton number density $n_{\bf 4}= (m_e/m_p)n_{l,\bf 4}$ (all number densities are given in comoving frames).
 
Dynamical evolution of the jet is described in detail  in Section \ref{s4-4}. Here we use numerical values 
of parameters presented there for fiducal values of determinative parameters: the  spin down energy
$E_{\textrm{sd}}=10^{52}$ erg,  the  ISM (stellar wind)  number density
$n_w = n_{\bf 1} = 10^{-3}$ cm$^{-3}$, the Lorentz factor of
magnetar wind jet $\gamma_\textrm{j}=10^3$, the jet duration
$t_\textrm{j}=t_{\textrm{sd}}=10^4$ s, the isotropic correction 
$\omega=10^{-3}$ (all values  are given  in  the centre of explosion  (COE) frame), 
the fraction of magnetic energy in MWN $\epsilon_B=10^{-3}$, and the  comoving magnetic
field in MWN $B_{\bf 3}=2\times 10^{-2}$ G. 
 
Reverse shock reaches the inner boundary of the jet-like magnetar wind with  fiducal parameters
penetrates at a distance from COE $R_{\Delta}=6.5\times 10^{18}$ cm  
at a moment of time $t_{\Delta}=R_{\Delta}/c=3.2\times 10^{8}$ s.
For this time, the all magnetar wind leptonic
plasma has crossed TS and became the MWN plasma. Both swept up and shocked 
by external shock ISM (region {\bf 2})
and separated by contact discontinuity MWN (region {\bf 3}) move 
with the Lorentz factor $\gamma_{{\bf 2}}=\gamma_{\bf 3}\sim f^{1/4}
\gamma^{1/2}_\textrm{j}/\sqrt{2} \sim 5.8\times 10^1$.  
On the other hand, the Lorentz factor of the MWN plasma with respect 
to the undisturbed magnetar wind is 
${\bar\gamma_{\bf 3}}= f^{-1/4} \gamma^{1/2}_\textrm{j}/\sqrt{2}\sim 8.6$, 
that corresponds 
to the TS Lorentz factor $\gamma_{\textrm{TS}}=\sqrt{2}{\bar\gamma_{\bf 3}}\sim 1.2\times 10^1$.
All process of MWN formation up to the moment when all   magnetar wind lepton 
plasma crosses the TS and fill in MWN  is accompanied by an  electron-positron CR 
acceleration at TS. Similarly to the ordinary PWN case, practically all magnetar
wind kinetic energy is tranformed to accelerated particles with the post-shock comoving 
energy density $w=w_e+w_B$ as a sum of the electron $w_e=\epsilon_e w$ and the
magnetic field $w_B=\epsilon_B w$ energy density, where 
$\epsilon_e\simeq 1$ and  $\epsilon_B\ll 1$ are the fractions  of electron and 
magnetic field  energy density, correspondingly.  In our case the  energy densities in 
regions {\bf 3} and {\bf 2} are equal and 
$w_{\bf 3}=w_{e,\bf 3}+ w_{B,\bf 3} = w_{\bf 2}
=4\gamma_{\bf 2}n_{\bf 1}m_pc^2\sim 2.0\times 10^{-2}$ erg cm$^{-3}$   
with a small admix of the magnetic field energy density $\epsilon_{B,\bf 3}\sim 10^{-3}$
for $B_{\bf 3}=2\times 10^{-2}$ G.

At the end of MWN formation the comoving  radial size of MWN was of 
$\Delta_{\bf 3}\sim 8.5\times 10^{15}$ cm, the transverse radius of
$R_{\Delta}\theta_\textrm{j}\sim 4\times 10^{17}$ cm, the comoving volume
$V_{\bf 3}=\pi \Delta (R_{\Delta}\theta_\textrm{j})^2\sim 4.3\times 10^{51}$ cm, 
and the comoving age 
$t_{\bf 3}\sim 3\Delta_{\bf 3}/c\sim 8.5\times 10^{5}$ s. Maximum energy of 
accelerated CRs is determined by balance of the acceleration and synchrotron 
energy loss times and is equal to 
$E_{\textrm{cr},e,\textrm{max}} =4.4\times 10^{14}$ eV. 
Energy spectrum of accelerated CR is expected to be analogous to the mentioned
above typical PWN case, i.e., exponential cut-off broken power-law one (\ref{eq:23}).
Minimum energy of electrons that radiate in fast cooling regime when the energy 
loss time is equal to  the comoving age is 
$E_{\textrm{cr},e,\textrm{fc}}  \sim 1$ TeV. 
Total MWN energy reserve in the comoving frame 
$W_{\bf{3},\textrm{tot}} \sim w_{\bf 3}V_{\bf 3} \sim 8.6\times 10^{49}$ erg 
is similar
to the energy reserve of shocked ISM (region {\bf 2}). In the COE frame the total 
energy of regions {\bf 2} and {\bf 3} are also similar and equal to 
$W^{\textrm{obs}}_{\bf 3}\sim W^{\textrm{obs}}_{\bf 2}
\sim \gamma_{\bf 3}W_{\bf{3},\textrm{tot}} \sim E_{\textrm{rot}}/4$
in each of the two jets, as expected from the energy conservation law. 

To summarise, as it looks  for observer in the  COE frame,  about $1.3\times 10^{7}$ s
after launch the jet-like magnetar wind pulse with the duration of $\sim 10^4$s, traveling 
the distance $\sim 4\times 10^{17}$ cm,  is completely decelerated due to 
interaction with ISM, transferring its energy to the two-layer 
structure: shocked and accelerated by external shock ISM and 
shocked and decelerated by reverse TS magnetar wind plasma,
i.e., MWN. In each jet MWN consist of accelerated leptonic CRs 
with  the comoving energy reserve 
$\sim E_{\textrm{rot}}/4\gamma_{\bf 3}\sim 4\times 10^{49}$ erg 
and as a whole moves with the  Lorentz factor $\gamma_{\bf 3}\sim 57$. Layer of  
accelerated  ISM has similar energy and Lorentz factor. 
After the mentioned above time, when the  reverse shock reaches the inner boundary of the jet,
the magnetar wind pressure vanishes and MWN enters the stage of free expansion in vacuum and 
CRs of MWN freely spread without adiabatic losses and diffusively feel a magnetar outskirts.  
Their mean energy increases in $\sim \gamma_{\bf 3}\sim 57$ 
times in the COE frame. Therefore 
we can expect now in magnetar outskirts presence of these MWN-born leptonic 
CR bubble with a typical for PWN ECBPL spectrum, total energy 
$\sim 0.5 E_{\textrm{rot}}$ from the both jets and 
a radius of $\sim 40$ pc according to (\ref{eq:22}). We can compare these predictions
with parameters of CRs needed for explanation of $\gamma$-ray emission from SGR1900+14 outskirts in the  MWN model. 

It is important to note that obtained above estimates of CR parameters
are rather upper limits because they correspond to the minimum energy losses
of leptonic CRs in MWN.

About a half of energy, deposited by jets in the ISM and stored in  thermal
and kinetic energy of mentioned above layers of shocked ISM, will support 
further a collimated jet-like baryonic plasma flow with a decreasing Lorentz 
factor $\gamma_{\bf 2}$ untill
$\gamma_{\bf 2}\theta_\textrm{j}\gtrsim 1$, and after that the jet  losses a collimation and the flow starts to 
spherize and at a distance of order of $l_{\textrm{Sed}}\sim 2\times 10^{20}$ cm  enters the Sedov stage 
according to Eq. (\ref{eq:8})--(\ref{eq:9}). At present, the non-relativistic jet-created shock
(JCS) has parameters, similar to the previous HNR  case, but with the something smaller energy
$E_{\textrm{JCS}} \sim E_{\textrm{rot}}/2$ (contributions of the both jets are combined),
the shock radius $R_{\textrm{JCS}}\approx 30$ pc and the velocity 
$V_{\textrm{JCS}}\approx 6\times 10^3$ kms$^{-1}$. In this case we expect 
$W_{\textrm{cr},p}\sim 5\times 10^{50}\eta_{p,-1}$ erg in proton CRs and 
$W_{\textrm{cr},e}\sim 5\times 10^{48}\eta_{p,-1}K_{ep,-2}$ erg in electron CRs.   

It is worth note, that contrary to the classical PWN scenario 
of HE-VHE emission generation, 
where only accelerated leptonic CRs produce $\gamma$-ray emission,
in our case of the  MWN scenario
together with accelerated in MWN leptonic CRs we inevitably 
get also proton and electron CRs 
accelerated at an external shock created by magnetar wind jet.
Therefore, the leptonic scenario in the MWN 
case includes contributions from the two lepton populations  -- of the  MWN leptons 
(electron-positron pairs) and  of the  external shock accelerated ISM  electrons (Eq.\ref{eq:24}).

To summarise,  in the PWN  model the observational data can be explained 
only when the total energy of radiating electrons is 
$W_{e,\textrm{tot}}\approx (1 - 5)\times 10^{50}d_{\textrm{10kpc}}^{2} \textrm{ ergs}$. 
Such energy considerably exceeds the expected CR energy in the case of ordinary
distant PWNe, but it can be provided by a newborn magnetar with a  millisecond
period of rotation (Table~\ref{parameters}).
Like the case of HNR model,  for the magnetic  field inside the stellar wind bubble
$B_w\lesssim 2\mu$G the synchrotron emission  does not disturb radio band upper limits
(Fig.~\ref{fig:PWN_BPL_IC}-\ref{fig:PWN_TwoPopulations_IC}). 

\subsection{Source confusion}\label{s4-4} 

For the $\gamma$-ray sources in the Galactic plane a source
confusion  is a notable problem, 
especially for extended sources. Among the  47 unidentified HGPS sources 36 have 
signatures of source confusion \citep{2021A&A...647A..68D} 

One of the source confusion reason in the SGR 1900+14 region may be
connected  with TeV halos. Recent observations of TeV halos
around some  middle-age pulsars  
\citep[][and references therein]{2017PhRvD..96j3016L,PhysRevD.100.043016}
can be explained as IC emission of high energy electrons and positrons, 
accelerated at the  pulsar wind
TSs,  but diffusively penetrated  outside the PWNe at the 
distances $\sim$ 10-30 pc and 
interacting with the background radiation fields.  TeV halos are expected to 
occur frequently, from  
10-50 TeV halos amount detected unidentified sources and  PWN candidates
up to sim 50–240 TeV halos
in future HAWC and CTA observations \citep{PhysRevD.100.043016}

In ATNF Pulsar Catalogue \citep{2005AJ....129.1993M} among of
middle-age (40-500 kyr) pulsars  in a circle of radius $0^{\circ}.5$ 
degree  around HESS J1907+089 position there are 
PSR J1907+0918  and  PSR J1908+0909,   whereas in a circle of radius $1^{\circ}$  there are
additionally PSR J1903+0925,   PSR J1908+0909, and   PSR J1909+0912, as well as  15 pulsars 
of unknown ages from The FAST Galactic Plane Pulsar  Snapshot survey
\citep{2021RAA....21..107H}. It is worth note that a full population of middle-age pulsars, 
including  “invisible” ones  with misaligned beaming angles, is about  four times  numerous
\citep{2017PhRvD..96j3016L}. It  is possible  that nearby  ($\lesssim 1$ kpc  in order to avoid an energy crisis)  invisible pulsars with TeV halos contribute to or even dominate in observed TeV flux.   
 
 $\gamma$-ray fluxes and spectra of detected  sources towards   the SGR 1900+14 region in 
 Fermi Large Area Telescope Fourth Source Catalog  \citep{2020ApJS..247...33A}, 
 especially the extended source 4FGL J1908.6+0915e   can be biased by the source confusion as well. 
 This extended source is overlapping with the  SGR 1900+14 and  the SNR G42.8+0.6. But,
 as  mentioned above, this SNR is detected only as a weak radio source 
 without any another non-thermal activity and is unlikely to be able to 
 generate a notable  $\gamma$-ray emission.  
There are  also three unidentified 4FGL sources in this region: J1908.7+0812, J1910.2+0904c, and 
J1911.0+0905 (sources 1, 2, and 3 in Fig.~\ref{fig:SGR_vicinity}, correspondingly). 
4FGL J1910.2+0904c has a flag 5 as an alert to confusion and  an append c 
(coincident with an interstellar clump) and  is very likely physically connected with
the SFR W49A \citep{Xin...Guo}. The SFR W49A includes a lot of young massive stars and HII regions embedded in a dense molecular cloud with promising conditions for $\gamma$-ray emission of stellar wind shock accelerated CRs, but without signatures of SNRs. Potentially, it could contribute to the $\gamma$-ray emission from a western hemisphere of an extended source 4FGL J1908.6+0915e (Fig. 3 in \cite{2017ApJ...835...30L}).  4FGL J1911.0+0905 is associated with the SNR W49B (G043.3-00.2), known as the TeV source HESS J1911+090 \citep{2018A&A...612A...1H}, but  due to a point-like type and considerable spacing we do not expect a severe source confusion with the SGR 1900+14 region.

\section{Comprehensive analysis of the Hypernova model}\label{s5}

We have shown that gross characteristics of the observed HE and VHE $\gamma$-ray emission from the region of the 
SGR 1900+14 together with the absence of detectable emission in lower energy bands  can be
explained in the framework of the Hypernova-like explosion of a SGR 1900+14 progenitor massive star
with the  dominant contribution of the newborn fast rotating magnetar in the SN energy reserve.
The main requirement for a successful model is a high amount of an energy deposited in accelerated
hadrons and leptons in both cases of  the $\gamma$-ray emission generation: by the dominant
contribution of the hadronic mechanism in the HNR model (the necessary total energy of the
hadronic CR component is $W_{\textrm{cr},p}\sim 5\times 10^{50} \textrm{ erg}$) or by the  leptonic
mechanism in the MWN model ($W_{\textrm{cr},e}\sim 5\times 10^{50} \textrm{ erg}$) (Table~\ref{parameters}).   

There are observational and theoretical arguments supporting an evolution 
scenario with the SGR 1900+14 progenitor as a  Hypernova/Superluminous SN (SLSN).

\subsection{Hypernova signatures in the optical band}\label{s5-1}

The magnetar SGR 1900+14  agrees  in a sky position, a distance, and an age with the 4 BCE "po star" -- 
the visible by naked eye ($\sim 5$ mag) transient ($\sim 1$ month) immovable 
source -- from ancient Chinese  record dated 4 BC Apr 24
\citep{2002ApJ...569L..43W, 2005JHA....36..217S}. Indeed, for the  mentioned above 
distance $d_{\textrm{mag}} = 12.5\pm 1.7 \textrm{ kpc}$ and the visual extinction
$A_V=12.9 \pm 0.5$ mag \citep{2009ApJ...707..844D} this explosion 
would have been optically observable only when
the  absolute  magnitude of  Hypernova  near to the  maximum  light was
$M=m +5 - 5\log(d_{\textrm{mag}} / \textrm{pc}) - A_V \approx -23$ mag. 
Estimated  value of $M$ is slightly more than  in the case of the most 
luminous  SN  yet  found ASASSN-15lh (SN 2015L) with 
$M_{\textrm{AB}}=-23.5\pm 0.1$ mag \citep{2016Sci...351..257D}
and is similar to $M \gtrsim -23$ mag SLSNe cases \citep{2018SSRv..214...59M, 2019MNRAS.487.2215A}. 
Even for a more conservative estimate $A_V = N_{\textrm{H}}/((2.04\pm 0.05)\times 10^{21}$ cm$^{-2}$) for 
the first Galactic quadrant \citep{2017MNRAS.471.3494Z}, so that $A_V=9.3$ mag from X-ray band 
suggested value $N_\textrm{H} = 1.9\times 10^{22}$ cm$^{-2}$, 
the absolute magnitude of the SGR 1900+14 progenitor $M \approx -19.4$ mag 
is noticeably lower than the typical value for core collapse SNe. 
An improved SGR 1900+14 age estimate $\tau_{\textrm{c}}\approx 2.4$ kyr from new $P$ and $\dot P$ data 
\citep{2019PASJ...71...90T} agrees with the "po star" age, taking into 
account an uncertainty of 
spin-down rate $\dot P$   determination \citep{2019PASJ...71...90T}.

\subsection{Hypernova remnant formation} \label{s5-2} 

The second argument, supporting Hypernova nature of the SGR 
1900+14 progenitor, is an absence of SNR signatures in radio 
to X-ray band. \cite{2000ApJ...533L..17V}   have classified 
this cluster as embedded
in parental molecular cloud, but \cite{2013A&A...560A..76M}
did not find  signatures of a molecular
cloud (clump) as a dominated by mass component
of the star cluster Cl1900+14/SGR 1900+14 (see subection 2.2 for details).

Molecular clumps are destroyed by a newly
formed star backreaction  even before a SN era 
\cite[see discussion in][]{2019ARA&A..57..227K,2020MNRAS.tmp.2473D}.
\cite{2018ApJS..237...13L}
showed that rotating ($V=300 \textrm{ km} \textrm{ s}^{-1}$) massive ($M=15 {\textrm{M}}_{\odot}$) 
star with solar metallicity ([Fe/H]=0)\footnote{As we mentioned above, such a model provides an 
observed dust mass $M_{\textrm{dust}}\gtrsim 2{\textrm{M}}_{\odot}$ in the SGR 1900+14 IR shell.} evolves
to a  Wolf–Rayet (WR) evolution stage  with a  lifetime  
$t_{\textrm{WR}}\approx 3.68\times 10^{5}$ yr 
and  a total WR wind mass loss 
$M_{\textrm{w,He}}\approx 4.0 {\textrm{M}}_{\odot}$ of helium.

At the WR stage, for  the  typical WR wind with the  mass-loss rate  
$\dot M_{\textrm{w}}\sim  10^{-5}{\textrm{M}}_{\odot} \textrm{ yr}^{-1}$, 
the wind velocity $V_{\textrm{w}}\sim 10^{8} \textrm{ cm} \textrm{ s}^{-1}$ 
and the  mechanical luminosity
$L_{\textrm{w}} = 0.5{\dot M}_{\textrm{w}} V_{\textrm{w}}^2$ the expected  WR 
wind bubble radius $R_{\textrm{bub}}$  at  the time $t_w$ since the beginning
 of WR stage is \citep{1977ApJ...218..377W}:
\begin{equation} \label{eq:18}
    \begin{split}
        R_{\textrm{bub}}= \left( \frac{125}{154\pi}\frac{L_{\textrm{w}}}{\rho_{\textrm{cl}}} \right)
        ^{1/5}t_{\textrm{w}}^{3/5} \\
        = 8.6\times 10^{-2} {\dot M}_{\textrm{w,-5}}^{1/5} V_{\textrm{w,8}}^{2/5}n_{\textrm{cl,4}}^{-1/5}
        t_{\textrm{w,kyr}}^{3/5} \textrm{ pc}.
    \end{split}
\end{equation}
\noindent
In our case the  WR wind destroys parental molecular cloud when  
$R_{\textrm{bub}} \approx R_{\textrm{cl}} \approx 0.7 \textrm{ pc}$ at time moment 
$t_{\textrm{w}} = 33 \textrm{ kyr}$, that is,  shorter of  the WR stage lifetime 
$t_{\textrm{w}} << t_{\textrm{WR}}=368 \textrm{ kyr}$.
It means that the WR wind alone can destroy a molecular clump in $\sim 30$ kyr, 
even before the first SN explosion. Till the moment of the
first SN outburst, joint action of stellar winds of three $M\sim 14-18$ 
M$_{\odot}$ stars in the newborn SGR 1900+14 star cluster can blow-up  the gas 
component of the parent $M_{\textrm{cl}}\sim 10^3 {\textrm{M}}_{\odot}$ clump   
up to $\sim 40 \textrm{ pc}$, forming an extended 
($\sim 10-20 \textrm{ pc}$) shell-like halo of fragmented debris of wind bubble
shells with the mean number density $n_{\textrm{sh}}\sim 10 \textrm{ cm}^{-3}$
and the total cluster plus ISM ($n_{\textrm{ISM}}\sim 1 \textrm{ cm}^{-3}$) swept
mass $\sim 10^4 {\textrm{M}}_{\odot}$
\citep{2012A&A...547A...3V, 2015MNRAS.448.3248G, 2019MNRAS.tmp.2162H,2020MNRAS.tmp.2473D}. 

In the Hypernova case, the gas component of ejecta ( without of 2 M$_{\odot}$ of dust slell) 
of $M_{\textrm{ej}}\sim 3 {\textrm{M}}_{\odot}$ will be  accelerated by a newborn 
magnetar with a rotation energy $E_{\textrm{rot}}\sim 10^{52}$ erg up to the velocity
$V_{\textrm{ej}}\sim (2E_{\textrm{rot}}/M_{\textrm{ej}})^{1/2}$ 
$\sim 1.9\times 10^{9} \textrm{ cm} \textrm{ s}^{-1}$ 
\citep{2010ApJ...719L.204W, 2019ApJ...880..150S}. Accelerated ejecta will expand in a mix of
$n_{\textrm{w}}\sim 10^{-3} \textrm{ cm}^{-3}$ winds from the exploded WR-star 
and the still present in the star cluster 
two RSG and a few BSG  stars \citep{2009ApJ...707..844D}.  For a constant ejecta 
density profile (with power-law index
$n=0$) and a constant density of  bubble interior (with power-law index $s=0$),
HNR enters the ST stage at the distance 
$R_{\textrm{ST}}^{*} \approx 32 \textrm{ pc}$
at the moment of time
$t_{\textrm{ST}}^{*} \approx 1.6 \textrm{ kyr}$ and reaches 
in the ST regime $R_{\textrm{ST}}\approx 35 \textrm{ pc}$ at the present time 
$t_{\textrm{SNR}} \approx 2 \textrm{ kyr}$.

\subsection{ Infrared emission of the clumpy Hypernova ejecta}\label{s5-3}  

Acceleration of an ejecta shell by a magnetar wind is accompanied by the development of the Rayleigh-Taylor
instability at a wind bubble-ejecta boundary, which can 
considerably complicate the process of shell acceleration
\citep{2017ApJ...845..139B}. Recent 3D-modelling of an initial interaction of 
a powerful  MWN 
($E_{\textrm{MWN}}\approx E_{\textrm{rot}}\sim 10^{52} \textrm{ erg}$) with a typical
SN ejecta 
($E_{\textrm{ej}} \sim 10^{51} \textrm{ erg}$), presented in \cite{2019ApJ...880..150S}, 
shows that considerable wind 
energy dominance  results in the Rayleigh-Taylor instability  of the MWN surface.

In our case of $L_{\textrm{sd}}=10^{48}$ erg s$^{-1}$ and $t_{\textrm{sd}}=10^4$ s  MWN with
the time-dependent  radius  $R_{\textrm{MWN}}\approx 2\times 10^{11} (t/t_{\textrm{sd}})^{5/4}$ cm 
is deep inside the exploding star.

The relativistic MWN plasma mixes  with the SN ejecta and  destroys
a regular shell-like structure of the ejecta. 
Numerous relativistic jet-like MWN plasma flows penetrate a steep gradient of 
ejecta layers  and accelerate an external ejecta layers by  mildly relativistic
forward shock.

This escaping through ejecta channels plasma carries out a dominant part 
of a magnetar wind energy, living a dense clumpy part 
of an ejecta with a relatively low energy gain.    
In the case of considered in \cite{2019ApJ...880..150S} ejecta  of 
$M_{\textrm{ej}} = 10\,{\textrm{M}}_{\odot}$  this clumpy part of 
ejecta with a mass $M_{\textrm{cl}}\approx 0.5\, M_{\textrm{ej}}$  reaches a typical 
velocity of $\sim 0.01c$ and for the SNR age of 2 kyr 
will be at distance of $\sim 6 \textrm{ pc}$ from the outburst place. 
In our case of the $M_{\textrm{ej}} \approx 5\, {\textrm{M}}_{\odot}$ ejecta,
a fragmentation of the ejecta is expected earlier and with lower debris 
velocities in accordance with the  observed SGR 1900+14 IR shell radius of $\sim 2 \textrm{ pc}$. 

The discussed  in subection 2.2  elliptical fit to the IR ring with  ratio of axes $\approx 2:1$
\citep{2008Natur.453..626W,2017ApJ...837....9N}
We  propose another model
of IR morphology, where observed IR ring of radius 
$R_{\textrm{ring}}\approx 27^{''}\approx 1.7 \textrm{ pc}$ consists 
of a newly formed dust 
in the Hypernova ejecta and is centred 
on the position of the magnetar's birthplace about 2 kyr ago (Fig.~\ref{fig:IR_ring}).
The newborn magnetar with  the transverse velocity
$V_\textrm{t}=130$  km s$^{-1}$ needs just 2 kyr   for traveling 
the distance of $5^{''}$ or 0.3 pc  from  
 its birthplace to the present position.
In our model only  dust-free cavity in the inflated part of 
ring was produced by the {\it anisotropic} giant flare in August 1998 
(or during the possible  previous, even more energetic, giant flare).
A short flare duration ($\lesssim 1$ s) in comparison with the magnetar's 
rotational period ($P=5.2$ s) results in an anisotropic 
energy release with a collimation angle $\sim 1$ rad in an agreement with observations.
Observed  dust mass $M_{\textrm{dust,out}} \sim 2{\textrm{M}}_{\odot}$
can also be explained in star evolution models.
\cite{2019MNRAS.484.2587M} shown, that evolution of a rotating 
($V_{\textrm{rot}}=300$ km s$^{-1}$)
calibrated (fixed) energy star model with an initial mass 
$M_{\textrm{ini}}=15\,{\textrm{M}}_{\odot}$  and solar metallicity ([Fe/H]=0) 
ends by a pre-SN mass of
$M_{\textrm{preSN}}=6.22(6.22)\, {\textrm{M}}_{\odot}$ and a Ib SN outburst with
an explosion energy $E_{\textrm{SN}}=0.93 (1.2)\times 10^{51}$ erg, 
an ejected mass of $M_{\textrm{ej}}=4.23(5.21)\, {\textrm{M}}_{\odot}$ and 
dust mass of $M_{\textrm{dust}}=1.5(2.25)\, {\textrm{M}}_{\odot}$.
An additional energy supply from a newborn magnetar with a rotational energy 
$E_{\textrm{rot}} \approx 2\times 10^{52}P_{\textrm{i,-3}}^{-2} \textrm{ erg}$
will provide a Hypernova type of an explosion
\citep{2015ApJ...805...82M,2016ApJ...818...94K,2018ApJ...857...95M, 2019ApJ...871L..25P}. 

\subsection{Magnetar wind nebula formation in a collimated jet-like magnetar wind} \label{s5-4}

Relativistic outflow from a magnetar in the form of a collimated jet is expected in
magnetar-related Hypernovae with possible long GRBs 
\citep{2003ApJ...589..871A,2009PhRvD..79j3001M,2019ApJ...871L..25P}.
Hydrodynamics  of jet interaction with ISM is analogous to the case of GRB jet \citep{1999PhR...314..575P}.

Dynamical evolution of jet depends, first of all,  on jet's energy $E^{\textrm{iso}}$.
Similar to classical PWN case Poynting flux dominated jet at large distances 
transforms into kinetic energy dominated jet with $\sigma_\textrm{j}\lesssim 1$ and increased 
correspondingly Lorentz factor $\gamma_\textrm{j}\sim \gamma_{\textrm{j,ini}}\sigma_{\textrm{j,ini}}$: 
$E^{\textrm{iso}}\approx 4\pi r^2 n_{l,\bf 4}m_e c^3 \gamma^2_{\textrm{j}}t_\textrm{j}$.
Another important parameters are  the scale of non-relativistic (Sedov stage) shock expansion
\begin{equation} \label{lsed}
 l_{\textrm{Sed}}=\left(\frac{E^{\textrm{iso}}}{n_{\bf 1}m_pc^2}\right)^{1/3}
 = 1.8\times 10^{20} \left(\frac{E_{\textrm{rot,52}}}{\omega_{-3}n_{{\bf 1},-3}}\right)^{1/3} \textrm{cm},
\end{equation}
undisturbed jet's  length $\Delta=ct_\textrm{j}=3\times 10^{14}t_{\textrm{j,4}}$ cm,
lepton number density in jet at the distance $r$
\begin{equation} \label{nl4r}
n_{l,\bf 4}(r) = 
\frac{E_{\textrm{rot}}}{4\pi r^2\omega\Delta \gamma_\textrm{j}^2 m_ec^2}
= 3.3\times 10^{3} \frac{E_{\textrm{rot,52}}}
{\omega_{-3}t_{\textrm{j},4}\gamma^2_{\textrm{j},3}r^2_{18}} \textrm{cm}^{-3}
\end{equation}
the distance $R_{\gamma}= l_{\textrm{Sed}}\gamma_\textrm{j}^{2/3} $ at which
the jet  starts to slow down (loses half of the initial energy) 
and  the distance $R_{\Delta}$ at which reverse shock reaches the
inner boundary of jet (i.e. all
magnetar wind plasma penetrates the TS and formation of MWN is completed)
\begin{equation} \label{rdelta}
R_{\Delta}=(l_{\textrm{Sed}}^3\Delta)^{1/4} = 
6.5\times 10^{18} \left(\frac{E_{\textrm{rot,52}}t_{\textrm{j,4}}}
{\omega_{-3}n_{{\bf 1}, -3}}\right)^{1/4} \, \textrm{cm}.
\end{equation}
At that distance the jet's density to ISM density ratio is 
\begin{equation} \label{f41}
f=\frac{\rho_{\bf 4}}{\rho_{\bf 1}}= \frac{n_{\bf 4}}{n_{\bf 1}}=
4.2\times 10^1C_f\gamma_{\textrm{j},3}^{-2}, \, 
C_f=\left(\frac{E_{\textrm{rot,52}}}{\omega_{-3}
t^3_{\textrm{j},4}n_{{\bf 1},-3}}\right)^{1/2}. 
\end{equation}

With these parameters at hand, we can estimate the  physical parameters 
of newborn MWN at moment of time $t_{\Delta}=R_{\Delta}/c$ s.
Both parts of the two-component jet -- regions $\bf 2$ and $\bf 3$ are
separated by the contact discontinuity and move  with the same Lorentz factor
\begin{equation} \label{Gamma23}
\gamma_{\bf 2}=\gamma_{\bf 3} = f^{1/4} \gamma^{1/2}_\textrm{j}/\sqrt{2}
= 5.8\times 10^1 C_f^{1/4}.
\end{equation}
The Lorentz factor of MWN plasma with respect to the undisturbed magnetar wind is
\begin{equation} \label{Gammabar3}
{\bar\gamma}_{\bf 3}= f^{-1/4} \gamma^{1/2}_\textrm{j}/\sqrt{2} =
8.6 C_f^{-1/4}\gamma_{\textrm{j},3},
\end{equation}
that corresponds to the TS Lorentz factor
$\gamma_{\textrm{TS}}=\sqrt{2}{\bar\gamma_{\bf 3}}$ in the free wind frame.
The energy densities $w$ are the same in both regions
\begin{equation} \label{w23}
w_{\bf 2}=w_{\bf 3} = 4\gamma_{\bf 2}^2n_{\bf 1}m_p c^2
=2.0\times 10^{-2}n_{{\bf 1},-3}C_f\, \textrm{erg cm}^{-3}.  
\end{equation}
As in the case of ordinary PWN magnetic field energy density 
we parametrise by magnetic energy
fraction $\epsilon_B=w_B/w$ with $\epsilon_B\sim 10^{-3}$.
For the comoving magnetic field  in MWN we obtain
\begin{equation} \label{B3-1}
B_{\bf 3}=(8\pi \epsilon_B w_{\bf 3})^{1/2} =
2\times 10^{-2} \epsilon^{1/2}_{B,-3}n^{1/2}_{{\bf 1},-3}C_f^{1/4} \, \textrm{G}.
\end{equation}

The comoving size of MWN is 
\begin{equation} \label{B3-2}
\delta_{\bf 3} = \Delta \gamma_\textrm{j}/4{\bar \gamma}_{\bf 3}
=8.5\times 10^{15}t_{\textrm{j},4}C_f^{1/4}\, \textrm{cm}
\end{equation}
and the comoving  MWN formation time is 
$t_{\textrm{MWN}}= 3\delta_{\bf 3}/c=8.5\times 10^5t_{\textrm{j},4}C_f^{1/4}$ s. 
Maximum energy of accelerated leptonic CRs can be  determined from the equality
of the acceleration time and the energy loss time (Eq. (\ref{emax})  and in our case is
$E_{e,\textrm{max}}= 6.0 \times 10^{14} B^{-1/2}_{{\bf 3},-2}$ eV. 
At the same time, CRs with the energy
loss time smaller than the MWN age $t_{\textrm{MWN}}$ or with
$E_e\gtrsim 4.5\times 10^{12}t_{\textrm{j},4}^{-1}C^{-1/4}_f$ eV
will radiate in the  fast cooling regime.

\section{Discussion}\label{s6}

Above we have modelled observed $\gamma$-ray emission from the  SGR 1900+14 
region for both hadronic and leptonis scenarios from three classes of 
potential sources  presented or expected in the area under consideration: 
SNRs (a magnetar-connected SNR,  the SNR G042.6, and the SNR  W49B), 
PWNe (a hypotetical magnetar-connected PWN and TeV halos), and 
SFRs (Cl 1900+14 and W49A)

 All of them, although to varying degrees, can contribute to 
 the observed $\gamma$-ray emission. 
 But while there are several potential sources, none are without problems. 
 It is useful to compare this region with the region around the  mentioned 
 above magnetar SGR 1806-20 (of 8.7 kpc distance)  with similar GeV-TeV fluxes
 and morphology: the  extended  4FGL J1808.2-2028e  source  with the radius of 
 $\Theta_{\textrm{ HE}}=0^{\circ}.65$ \citep{2016ApJ...827...41Y} is associated 
 with  the unidentified point-like HESS J1808-204 source
 \citep{2018A&A...612A..11H,2018A&A...612A...1H}. 
 In addition to the magnetar SGR 1806-20 itself with a possible PWN, potential
 sources include its host: the massive star cluster Cl* 1806-20 with four WR
 stars and four OB supergiants with powerful stellar winds.  
 Once more,  the hypergiant LBV 1806-20 wind powers a radio-nebula G10.0-0.3.
 An another promising $\gamma$-ray source towards  the magnetar SGR 1806-20 region 
 is a shell-type SNR G9.7-0.0 (of 4.7 kpc distance) interacting with a nearby MC.
It is worth noting that  in the 3-500 GeV band the extended morphology of 
4FGL J1808.2-2028e consists of two separated clumps,  coincident with 
HESS J1808-204 and the SNR G9.7-0.0, respectively \citep[][and references
therein]{2016ApJ...827...41Y,2018A&A...612A..11H,2018A&A...612A...1H}

All these classes of potential sources are presented in the 
SGR 1900+14 region as well. 
But their  abilities  to provide the observed $\gamma$-ray fluxes do not look so 
optimistic. Magnetar's host massive star cluster Cl 1900+14 with only two M5 RSG 
stars and without OB or WR stars  (see Section \ref{s2-3} for details) cannot ensure  
effective CR acceleration. The other stellar cluster  - the luminous SFR W49A looks
more promising, but displays only a weak  point-like GeV $\gamma$-ray source 
(if any). Analogously to the  4FGL J1808.2-2028e case, the SFR W49A could 
contribute to one overlapping
him clump in a hypothetical two-clump structure of 4FGL J1908.6+0915e

Among the two SNRs in the SGR 1900+14 region the  SNR G42.8+0.6 
manifests himself only 
as a weak radio shell without any X-ray or $\gamma$-ray activity,
whereas  the SNR G043.3-00.2 (W49B) is detected as a point source in HE and VHE
$\gamma$-rays, but out of the region considered. 

In this situation we expect  that the dominant contribution to 
$\gamma$-ray emission from the 
SGR1900+14 region most likely is  provided by the magnetar-related SNR/MWN and 
mentioned above hypothetical  TeV halos due to the source confusion.

\subsection{Magnetar-connected HNR/MWN model}\label{s6-1}
 
The magnetar-related solution  with sources at the  magnetar distance 
of 12.5 kpc requires the Hypernova-type energy reserve 
$\sim 10^{50}-10^{51}$ erg in CRs 
in both hadronic and leptonic scenarios. In Section \ref{s4} we 
presented a detailed description 
and verified reliability of the HNR/MWN model. Hypernovae
and GRB remnants have been proposed
as unidentified TeV sources  for a long time
\citep{2010ApJ...709.1337I,2013ApJ...774...74F}.
A potential weakness of this model is the small number of Hypernovae in our Galaxy. 
The measurement of the local core-collapse SN (CCSN)  rate and  of the subclass of 
hydrogen-free superluminous SN  (SLSN-I) rate using SN rates from the Palomar 
Transient Factory  gives the local ratio of SLSN-I to  all  CCSN types of 
$\sim 1/3500^{+2800}_{- 720}$ \citep{2021MNRAS.500.5142F}. For  the Galactic CCSN 
rate $\sim 1/30$ yr$^{-1}$ and a    typical HNR age $\sim 10^5$ yr the expected 
number of HNRs in  our Galaxy is about a few.  In particular, HNRs may 
reside in the Galactic
center \citep{2020ApJ...891..179H} and in the Cygnus region
\citep{2021ApJ...919...93F}.

Generally, magnetar-connected HNR/MWN models give reasonable explanation of 
the set of observational multiwavelength  data concerning the magnetar 
SGR 1900+14 and its  neighbourhood. But some parameters  of these model are far 
from their typical values. So we see the main challenge of the HNR model in
the requirement of the high  value of total hadronic CR energy
$W_{\textrm{cr},p}\sim 5\times 10^{50}(n_{\textrm{sh}}/10 \textrm{ cm}^{-3})^{-1}$ erg 
(Table~\ref{parameters}, Figs.~\ref{fig:PionIC_PL}, \ref{fig:PionIC_ECPL}).     
  
As in the HNR case, the main challenge of the MWN model is the requirement of the high 
value of the total energy of accelerated leptons $W_{\textrm{cr},e}\sim 5\times 10^{50}$ erg,
that is, on the order of the total energy of protons in the HNR model  
(Table~\ref{parameters}, Figs.~ \ref{fig:PWN_BPL_IC}-\ref{fig:PWN_TwoPopulations_IC}).

A newborn millisecond magnetar can help to solve the energy problem, but another challenge 
in the  magnetar case  is the low probability of  magnetars' millisecond periods. Recent 
results concerning the  parameters of magnetars, connected with the long GRBs and the Type I SLSNe
suggest possible millisecond equilibrium spin period due to the fallback accretion \citep{2020ApJ...903L..24L}.

Some fine tuning is connected with the  magnetar jet and the jet-created MWN. For parameters considered 
leptons in the transient MWN radiate in the  slow cooling regime without serious energy losses and are
spread in the magnetar region with considerable part of jet energy. Therefore our
estimates of the energy reserve of accelerated CRs can be considered as a lower limit. 
  
Presented here model of the  magnetar jet does not take into account a possible
baryon contamination. Each baryon (proton) in the relativistic jet carries energy
$m_p/m_e$ times higher than electron. Therefore, even small admix of baryons
considerably reduces the leptonic energy reserve and, correspondingly, leptonic 
emission. Our estimations can be considered as an upper limit on expected 
leptonic emission for magnetar jets.    

Till now, only $\gamma$-ray emission is detected from the magnetar SGR 1900+14 outskirts.
In both HNR and MWN cases the radio and X-ray non-detection corresponds to a low value
of magnetic field $B~\lesssim~2~\mu\textrm{G}$  in the magnetar neighbourhood. Such
values of $B$ are expected in a large $\gtrsim  30$ pc  rarefied 
$n_{\textrm{w}}\sim 10^{-3} \textrm{ cm}^{-3}$ stellar wind bubbles. 

\subsection{Testing of alternative models}\label{s6-2}
 
As mentioned above, a contribution from a TeV halo around a nearby
(of $\sim 1$ kpc distance) invisible pulsar  could also be an option. 
The contribution of other sources, including SFRs,  cannot be ruled out
as well. Additional multiwavelength -- from radio to multi-TeV bands --  
and spectral observations are needed to clarify the
nature of the $\gamma$-ray sources in this region.

First of all, new observations in TeV band are very promising to establish the 
nature of sources. Already future observations of the SGR 1900+14 region
by the forthcoming Cherenkov Telescope Array (CTA, \cite{2019scta.book.....C}) 
in the short-term programme (years 1-2)  of the CTA Galactic Plane Survey by the 
South observatory with an equivalent exposure time 11.0 h ( a point-source sensitivity 
2.7 mCrab) should detect sources in the SGR 1900+14 region. The combination of 
high resolution imaging atmospheric Cherenkov telescopes (IACTs) H.E.S.S., CTA  
and wide field of view  air shower arrays (ASAs) HAWC, LHAASO in forthcoming
observations will reveal the spectral and morphological features of sources. 
In the hadronic  HNR model we expect a shell-like or an arch-like 
morphology and a power-law $\Gamma_{\textrm{cr,p}} = 2.2-2.5$ 
spectrum with an exponential cut-off, whereas both leptonic MWN and TeV halo models
are expected to have a filled center plerion-like morphology  and a hard sub-cutoff
spectrum. Once more, in the TeV halo case the spatial  extent of  associated 
SNR/X-ray PWN, if observed, is significantly larger/smaller than  of the halo himself
\citep{2017PhRvD..96j3016L,PhysRevD.100.043016}. 

The high angular  resolution of the CTA and \fermilat in multi-GeV band can help to 
investigate sub-structures and energy-depended morphology of the extended \fermilat 
4FGL J1908.6+0915e source. If the two-clump structure as in the case  
of the magnetar SGR 1806-20 is confirmed, a hybrid model 
of multi-GeV $\gamma$-ray emission may be more adequate. It may include mixes 
of hadronic and leptonic contributions from the clump that overlaps the SFR W49A 
and from the  magnetar - connected clump. Analysis of HE  spectra  of clumps 
can help to disentangle confused sources 
\citep{2016ApJ...827...41Y,2019scta.book.....C}.

Multi-wavelength observations in the radio to X-ray band with the high angular resolution  
open up additional possibilities for testing alternative models. New observations of
mentioned above pulsars in the SGR 1900+14 region by FAST
\citep{2021RAA....21..107H} will determine their  spin-down 
rate and thus their physical parameters as potential sources of  confused TeV halos.
It will also be important to look for radio-signatures  of synchrotron emission  
from PWNe and SN remnants as well as  for HI, CO, OH radio-lines due to a SNR 
shock-molecular cloud interaction \citep{2015SSRv..188..187S,2015ApJ...806...71L}. Detection of 
the shock-molecular cloud interaction would allow to clarify the distance to  the
$\gamma$-ray source, which is of fundamental importance for the HNR model. 

Promising perspective for clarifying the nature of the $\gamma$-ray sources in the 
SGR 1900+14 region  is connected with 
the  forthcoming extended ROentgen Survey with an Imaging Telescope Array (eROSITA)
on board the Spectrum-Roentgen-Gamma (SRG) mission, that will be about 25 times
more sensitive than the ROSAT All-Sky Survey in the soft X-ray (0.2 - 2.3 keV) band 
\citep{2021A&A...647A...1P}

\section{Conclusions}\label{s7}

We have modelled the observed HE and VHE $\gamma$-ray emission from the 
sky region of the  magnetar SGR 1900+14 as a result of 
the contributions of individual sources available here, namely, 
the SNRs G42.8+0.6 and  W49B,
the SFRs Cl 1900+14 and W49A, as well as an  undetected yet  
magnetar-connected SNR and possible MWN. Source confusion with TeV 
halos is also probable.

Our main conclusions can be summarised as follows.

1. Physical conditions in the SNRs G42.8+0.6 and  W49B, as well as in 
the SFRs Cl 1900+14 and W49A cannot provide a significant contribution to the 
observed HE and VHE $\gamma$-ray emission.
 
2. A set of observational multiwavelength  data may be explained in a model of a 
magnetar-connected SN, exploded about 2 kyr ago at a distance of 12.5 kpc 
and mentioned in ancient Chinese records on 4 BC Apr 24. SN should be 
of Hypernova type with the explosion energy $E_{\textrm{HNR}} \sim 10^{52}$ erg
derived from the rotational energy 
$E_{\textrm{rot}} \approx 2\times 10^{52}P_{\textrm{i,-3}}^{-2}$ erg  
of the newborn millisecond strongly magnetised  X-ray pulsar -- magnetar SGR 1900+14.
Equilibrium millisecond period of magnetar with
$B_{\textrm{s}}=4.3\times 10^{14}$ G is expected  due to the fallback accretion.

3. The extremely high initial magnetic dipole spin-down luminosity will result in 
energization of the Hypernova ejecta and in formation of the MWN.  
Relative shares of the energy input to the Hypernova ejecta and to the MWN are
model-dependent and can be estimated from the analysis of  new observations.

4. In the HNR model, the energy-dominated Hypernova ejecta with 
$E_{\textrm{HNR}}\approx  E_{\textrm{rot}} \approx 10^{52}$ erg 
evolves inside the low-density wind bubble creating the HNR 
at the ST stage at present.
Shock accelerated nuclei and electrons are diffusively spread over the observed
$\gamma$-ray sources  and $\gamma$-ray emission is generated mainly by the  hadronic 
mechanism ($pp$ collisions with subsequent neutral pion decay)  in the dense 
($n_{\textrm{sh}}\sim 10 \textrm{ cm}^{-3}$)  extended  shell-like halo of 
swept up mass of $\sim 3\times 10^4\,{\textrm{M}}_{\odot}$. In the TeV band
the contribution of the leptonic mechanism (IC scattering of background CMB, 
IR and SL photons by shock accelerated electrons) can be dominant. 

5. The powerful relativistic  collimated jet-like magnetar wind can effectively 
penetrate the Hypernova ejecta, producing twin GRB-like jets in a rarefied 
stellar wind bubble. Due to the leptonic jet deceleration, the 
two transient MWNe with 
$E_{\textrm{MWN}}\lesssim 0.5 E_{\textrm{rot}} \approx 10^{52}$ erg 
arise and  the jet's energy is  partially  transferred to TS/reconnection accelerated
electrons and positrons. After an jet shutdown the  accelerated leptons spread over the
observed $\gamma$-ray sources and $\gamma$-ray emission is generated by the leptonic
mechanism. 

Certainly, a more realistic model should include contributions from  both hadronic and 
leptonic models, but  sparse observational data available do not allow us to build a fully 
quantitative model for the $\gamma$-ray emission of the SGR 1900+14 environment. 

6. The source confusion  caused by expected TeV halos in the Galactic plane is
actual for the region of SGR 1900+14.

In the next few years  new multiwavelength  observations in radio (FAST), X-ray
(eROSITA), and $\gamma$-ray  (MAGIC, H.E.S.S., HAWC and LHAASO) bands 
could clarify  source characteristics and 
improve our understanding of the physical processes 
in SNRs/MWNe and the nature of the $\gamma$-ray emission from the region of the magnetar  SGR 1900+14.

\section*{Acknowledgements}\label{s8}
We are grateful to the anonymous reviewers for very attentive and helpful 
comments and suggestions 
that helped us significantly improve the quality of the manuscript.
The Spitzer IR image of the  SGR 1900+14 environment is shown by courtesy NASA/JPL-Caltech. 

{\bf Data   availability   statement:} All observational data used in the paper are publicly 
available in the cited works. The datasets generated and analyzed during the current study 
are available from the corresponding author on reasonable request.



\bibliographystyle{mnras}
\bibliography{references} 



\bsp	
\label{lastpage}
\end{document}